\documentclass[12pt]{article}
\usepackage{amsmath,amsthm}

\begin{document}

\begin{center}
{\Large{\bf osp(1,2)-COVARIANT LAGRANGIAN
\\
QUANTIZATION
\\
\medskip
OF IRREDUCIBLE MASSIVE GAUGE THEORIES}}
\\
\bigskip\bigskip
{\large{\sc B. Geyer}}\footnote{E-mail: geyer@rz.uni-leipzig.de}
\\
\smallskip
{\it Naturwissenschaftlich-Theoretisches Zentrum and}
\\
{\it Institut f\"ur Theoretische Physik, Universit\"at Leipzig},
\\
{\it Augustusplatz 10/11, 04109 Leipzig, Germany}
\\
\bigskip
{\large{\sc P. M. Lavrov}}
\\
\smallskip
{\it Tomsk State Pedagogical University}
\\
{\it 634041 Tomsk, Russia}
\\
\bigskip
{\large{\sc D. M\"ulsch}}
\\
\smallskip
{\it Wissenschaftszentrum Leipzig e.V.}
\\
{\it Goldschmidtstr. 26, 04103 Leipzig, Germany}
\\
\bigskip\medskip
{\small{\bf Abstract}}
\\
\end{center}

\begin{quotation}
\noindent {\small{The $osp(1,2)$-covariant Lagrangian quantization of general 
gauge theories is formulated which applies also to massive fields. The 
formalism generalizes the $Sp(2)$-covariant BLT approach and guarantees 
symplectic invariance of the quantized action. The dependence of the
generating functional of Green's functions on the choice of gauge in the 
massive case disappears in the limit $m \rightarrow 0$. 
Ward identities related to $osp(1,2)$ symmetry are derived. Massive gauge 
theories with closed algebra are studied as an example.}}
\end{quotation}


\setlength{\baselineskip}{0.6cm}

\bigskip\medskip
\begin{flushleft}
{\large{\bf I. INTRODUCTION}}
\end{flushleft}
\bigskip
Recently, a general method for quantizing gauge theories in the 
Lagrangian formalism has been proposed \cite{1,2,3}
which is based on {\it extended} BRST symmetry, i.e. simultaneous invariance 
under both BRST and antiBRST transformations. It is characterized by a quantum 
action functional $S = S(\phi^A, \phi_{A a}^*, \bar{\phi}_A)$ depending, 
besides on the dynamical fields 
$\phi^A = (A^i, B^{\alpha_0}, C^{\alpha_0 a_0})$, also on related external 
sources (or antifields) $\phi_{A a}^*$, $\bar{\phi}_A$, where 
$A^i$, $B^{\alpha_0}$ and $C^{\alpha_0 a_0}$ are the gauge, the auxiliary and 
the (anti)ghost fields respectively, and both $a$ and $a_0$ indicate members 
of $Sp(2)$ doublets. 
To guarantee their (anti)BRST symmetry the action $S$ (and the related gauge 
fixed extended action $S_{\rm ext}$) is required to satisfy two master 
equations which are generated by two nilpotent, anticommuting differential 
operators, $\bar{\Delta}^a$. The method applies to irreducible as 
well as reducible, complete gauge theories with either open or closed gauge 
algebra. (The condition of irreducibility requires the generators of the 
gauge transformations to be linearly independent at the stationary point of 
the classical action, and the condition of completeness requires the 
degeneracy of the Hessian of the classical action $S_{\rm cl}(A)$ to be 
solely due to its gauge invariance \cite{4,5}).

Although this formalism is seemingly manifest $Sp(2)$-covariant among the 
solutions of the master equations, despite being allowed by the above 
requirements, there are both $Sp(2)$-symmetric and $Sp(2)$-nonsymmetric ones.
The symmetric solutions may be singled out by the explicit requirement 
of invariance under $Sp(2)$ transformations by additional master equations 
whose generating differential operators
$\bar{\Delta}_\alpha$ $(\alpha = 0,+,-)$ are related to the generators of 
the symplectic group $Sp(2)$. The algebra of these operators may be chosen 
to obey the orthosymplectic (super)algebra $osp(1,2)$. (Actually, its even
part is the algebra $sl(2)$ generating the special linear transformations, 
but due to their isomorphism to the algebra $sp(2)$ we will speak about 
symplectic transformations.)
Moreover, if also {\it massive} fields should be considered to 
circumvent possible infrared singularities occuring in the 
process of subtracting ultraviolet divergences, without breaking the 
extended BRST symmetry, then this algebra appears necessarily. 
Let us also mention that the $osp(1,2)$ superalgebra is present in 
many problems in which $N = 1$ superconformal symmetry is involved; e.g. 
in the minimal $N = 1$ superconformal models this symmetry appears in the
light-cone approach to two-dimensional supergravity \cite{6}. It is also
of interest in this respect that topological $osp(1,2)/osp(1,2)$ coset
theories can be used to describe the non-critical Ramond-Neveu-Schwarz
superstrings \cite{7}.
 
The goal of the present paper will be to generalize the BLT quantization 
procedure to another one being $osp(1,2)$-covariant. For the sake of 
simplicity we restrict ourselves to the case of irreducible (or zero-stage) 
complete gauge theories and thereby we follow very closely the exposition of 
Ref. \cite{1} 
(The extension to $L$-stage reducible theories will be dealt with in a 
succeding paper). We also used the condensed notation introduced 
by DeWitt \cite{8} and conventions adopted in Ref. \cite{1}; 
if not otherwise specified, derivatives with respect to the antifields are the 
(usual) left ones and that with respect to the fields are {\it right} ones. 
Left derivatives with respect to the fields are labeled by the 
subscript $L$, for example, $\delta_L/\delta \phi^A$ denotes the left 
derivative with respect to the fields $\phi^A$.
 
The paper is organized as follows. In Section II we shortly review the 
standard $Sp(2)$-covariant approach and point out how it will be generalized 
to the $osp(1,2)$-covariant quantization procedure. As a consequence of the 
enlarged algebra a canonical definition of the ghost number (Faddeev-Popov 
charge) is obtained. Furthermore, to be able to express this algebra through 
{\it operator identities} and to get nontrivial solutions of the generating
equations it is necessary to enlarge the set of antifields. 
In Section III the explicit construction of generating differential operators 
fulfilling the $osp(1,2)$ algebra is outlined, starting with the approximation 
of the action $S_m$ at lowest order in $\hbar$ which is assumed to be linear 
with respect to the antifields. In Section IV the gauge dependence of the 
generating functional of Green functions is studied and corresponding Ward 
identities are derived. It is shown that the mass terms destroy gauge 
independence of the $S$-matrix. In Section V we consider massive theories 
with closed gauge algebra, thereby extending the solution given in \cite{1}. 
In Section VI, restricting the fore-\break going result to the case $m = 0$, 
we construct a solution of the quantum master equation in the $Sp(2)$-approach
which is not $Sp(2)$-symmetric.
\bigskip\medskip
\begin{flushleft}
{\large{\bf II. GENERAL STRUCTURE OF osp(1,2)-COVARIANT
\\
$\phantom{\bf II.}$ QUANTIZATION OF IRREDUCIBLE GAUGE THEORIES}}
\end{flushleft}
\bigskip
Let us consider a set of gauge (as well as matter) fields $A^i$ with Grassmann 
parity $\epsilon(A^i) = \epsilon_i$ for which the classical action 
$S_{\rm cl}(A)$ is assumed to be invariant under the gauge 
transformations\footnote{
In the following an additional label $0$ is put on the indices $\alpha_0$ 
and $a_0$ to prepare the notation for later generalizations to reducible gauge 
theories.} 
\begin{equation}
\delta A^i = R^i_{\alpha_0} \xi^{\alpha_0},
\qquad
S_{{\rm cl}, i} R^i_{\alpha_0} = 0,
\end{equation}
where $\xi^{\alpha_0}$ and $R^i_{\alpha_0}$ are the (infinitesimal) gauge 
parameters and the gauge generators having Grassmann parity 
$\epsilon(\xi^{\alpha_0}) = \epsilon_{\alpha_0}$ and  
$\epsilon(R^i_{\alpha_0}) = \epsilon_i + \epsilon_{\alpha_0}$, respectively;
by definition $X,_j = (\delta/ \delta A^j) X$. 
We assume the set of generators $R^i_{\alpha_0}(A)$ to be linearly independent 
and complete. The (open) algebra of generators has the general form 
\cite{1}: 
\begin{equation}
R_{\alpha_0, j}^i R_{\beta_0}^j -
(-1)^{\epsilon_{\alpha_0} \epsilon_{\beta_0}} 
R_{\beta_0, j}^i R_{\alpha_0}^j =
- R_{\gamma_0}^i F_{\alpha_0 \beta_0}^{\gamma_0} - 
M^{ij}_{\alpha_0 \beta_0} S_{{\rm cl}, j},
\end{equation}
where $F_{\alpha_0 \beta_0}^{\gamma_0}(A)$, in general, are field dependent 
structure functions and $M^{ij}_{\alpha_0 \beta_0}(A)$ obeys the graded 
antisymmetry conditions
\begin{equation}
M^{ij}_{\alpha_0 \beta_0} = - (-1)^{\epsilon_i \epsilon_j} 
M^{ji}_{\alpha_0 \beta_0} = - (-1)^{\epsilon_{\alpha_0} \epsilon_{\beta_0}} 
M^{ij}_{\beta_0 \alpha_0}.
\end{equation}
Theories whose generators satisfy eqs. (2) and (3) are called
{\it general gauge theories}. In the case $M_{\alpha_0 \beta_0}^{ij} = 0$ 
the algebra is closed. 

The total configuration space of fields $\phi^A$ and their Grassmann parities 
are 
\begin{equation*}
\phi^A = (A^i, B^{\alpha_0}, C^{\alpha_0 a_0}), 
\qquad
\epsilon(\phi^A) \equiv \epsilon_A = 
(\epsilon_i, \epsilon_{\alpha_0}, \epsilon_{\alpha_0} + 1); 
\end{equation*}
here, the auxiliary fields $B^{\alpha_0}$ are $Sp(2)$-scalar whereas the 
(anti)ghosts $C^{\alpha_0 a_0}$ transform as a $Sp(2)$-doublet. Moreover, for 
each field $\phi^A$ one introduces two sets of antifields, a $Sp(2)$-doublet 
and a $Sp(2)$-singlet: 
\begin{equation*}
\phi^*_{A a} = (A^*_{i a}, B^*_{\alpha_0 a}, C^*_{\alpha_0 a a_0}), 
\qquad\!
\epsilon(\phi^*_{A a}) = \epsilon_A + 1,
\qquad\!
\bar{\phi}_A = (\bar{A}_i, \bar{B}_{\alpha_0}, \bar{C}_{\alpha_0 a_0}), 
\qquad\!
\epsilon(\bar{\phi}_A) = \epsilon_A.
\end{equation*}
Raising and lowering of $Sp(2)$-indices is obtained by the invariant tensor 
of the group,
\begin{equation*}
\epsilon^{ab} = \begin{pmatrix} 0 & 1 \\ -1 & 0 \end{pmatrix},
\qquad
\epsilon^{ac} \epsilon_{cb} = \delta^a_b.
\end{equation*}
Let us point to the fact that in the BLT approach the internal $Sp(2)$ index 
$a_0$ of the third component is a dummy one, i.e. it is not affected by 
operations being introduced by the present approach.

Let us now shortly review the $Sp(2)$-covariant BLT quantization scheme. The 
basic 
object is the bosonic action $S = S(\phi^A, \phi^*_{A a}, \bar{\phi}_A)$ 
satisfying the extended quantum master equations (i.e. the generating 
equations with respect to the extended BRST symmetry) 
\begin{equation}
\bar{\Delta}^a\, {\rm exp}\{ (i/ \hbar) S \} = 0,
\qquad
\bar{\Delta}^a = \Delta^a + (i/ \hbar) V^a,
\end{equation}
where the odd (second-order) differential operators $\bar{\Delta}^a$ possess 
the important properties of nilpotency and (relative) anticommutativity:
\begin{equation}
\{ \bar{\Delta}^a, \bar{\Delta}^b \} = 0.
\end{equation} 
The solution of (4) is sought as a power series in Planck's constant $\hbar$:
\begin{equation*}
S = \sum_{n = 0}^\infty \hbar^n S_{(n)},
\end{equation*}
with the boundary condition 
$S|_{\phi^*_a = \bar{\phi} = \hbar = 0} = S_{\rm cl}(A)$.

To remove the degeneracy of the action $S$ a gauge has to be introduced with 
the following properties: First, it should lift the degeneracy in 
$\phi^A$ and, second, it should retain eqs. (4), thereby providing the 
extended BRST symmetry also for the gauge fixed action denoted by 
$S_{\rm ext} = S_{\rm ext}(\phi^A, \phi^*_{A a}, \bar{\phi}_A)$. Introducing 
the gauge-fixing $Sp(2)$-invariant bosonic functional $F = F(\phi^A)$ the 
action $S_{\rm ext}$ is defined by  
\begin{equation*}
{\rm exp}\{ (i/ \hbar) S_{\rm ext} \} = 
\hat{U}(F) \,{\rm exp}\{ (i/ \hbar) S \}, 
\end{equation*}
where the operator $\hat{U}(F)$ has the general form
\begin{equation*}
\hat{U}(F) = {\rm exp}\{ (\hbar/ i) \hat{T}(F) \},
\qquad
\hat{T}(F) = \hbox{$\frac{1}{2}$} \epsilon_{ab} 
\{ \bar{\Delta}^b, [ \bar{\Delta}^a, F ] \}.
\end{equation*}
Here, $\hat{T}(F)$ has been choosen such that, by virtue of (5), it commutes 
with $\bar{\Delta}^a$,
\begin{equation*}
[ \bar{\Delta}^a, \hat{T}(F) ] = 0;
\end{equation*}
hence $S_{\rm ext}$ obeys also eqs. (4):
\begin{equation*}
\bar{\Delta}^a\, {\rm exp}\{ (i/ \hbar) S_{\rm ext} \} = 0.
\end{equation*}

Let us now shortly state the essential modifications of the BLT formalism
to obtain the $osp(1,2)$-covariant quantization of an irreducible complete
theory of massive fields whose action $S_m$ depends on the mass $m$ 
as a further parameter. First, in addition to the $m$-{\it extended} 
generalized quantum master equations
\begin{equation}
\bar{\Delta}_m^a\, {\rm exp}\{ (i/ \hbar) S_m \} = 0,
\qquad
\bar{\Delta}_m^a = \Delta^a + (i/ \hbar) V_m^a,
\end{equation}
which ensure (anti)BRST invariance, the action $S_m$ is required to satisfy
the generating equation of $Sp(2)$-invariance, too:
\begin{equation}
\bar{\Delta}_\alpha\, {\rm exp}\{ (i/ \hbar) S_m \} = 0,
\qquad
\bar{\Delta}_\alpha = \Delta_\alpha + (i/ \hbar) V_\alpha,
\end{equation}
where $\bar{\Delta}_m^a$ and $\bar{\Delta}_\alpha$ are odd and even 
(second-order) differential operators, respectively (for explicit expressions 
see Section III below). 

As long as $m \neq 0$ the operators $\bar{\Delta}_m^a$ are neither nilpotent 
nor do they anticommute among themselves; instead, together with the operators 
$\bar{\Delta}_\alpha$ they generate a superalgebra isomorphic to $osp(1,2)$ 
(see Appendix A):
\begin{align}
[ \bar{\Delta}_\alpha, \bar{\Delta}_\beta ] &= (i/ \hbar)
\epsilon_{\alpha\beta}^{~~~\!\gamma} \bar{\Delta}_\gamma, 
\\ 
[ \bar{\Delta}_\alpha, \bar{\Delta}_m^a ] &= (i/ \hbar) 
\bar{\Delta}_m^b (\sigma_\alpha)_b^{~a}, 
\\ 
\{ \bar{\Delta}_m^a, \bar{\Delta}_m^b \} &= - (i/ \hbar) m^2 
(\sigma^\alpha)^{ab} \bar{\Delta}_\alpha, 
\end{align}
where
\begin{equation*}
(\sigma_\alpha)^{ab} = \epsilon^{ac} (\sigma_\alpha)_c^{~b} =
(\sigma_\alpha)^a_{~c} \, \epsilon^{cb} =
\epsilon^{ac} (\sigma_\alpha)_{cd} \, \epsilon^{db},
\qquad
(\sigma_\alpha)_a^{~b} = - (\sigma_\alpha)^b_{~a};
\end{equation*} 
here, the matrices $\sigma_\alpha$ ($\alpha = 0,+,-$) generate the group of 
special linear transformations:
\begin{gather*}
\sigma_\alpha \sigma_\beta = g_{\alpha\beta} + 
\hbox{$\frac{1}{2}$} \epsilon_{\alpha\beta\gamma} \sigma^\gamma,
\qquad
\sigma^\alpha = g^{\alpha\beta} \sigma_\beta,
\qquad
\hbox{Tr}(\sigma_\alpha \sigma_\beta) = 2 g_{\alpha\beta},
\\
g^{\alpha\beta} = \begin{pmatrix} 1 & 0 & 0 \\  0 & 0 & 2 \\ 0 & 2 & 0 
\end{pmatrix},
\qquad
g^{\alpha\gamma} g_{\gamma\beta} = \delta^\alpha_\beta,
\end{gather*}
$\epsilon_{\alpha\beta\gamma}$ being the antisymmetric tensor, 
$\epsilon_{0+-} = 1$. Notice, that $sl(2, R)$, the even part of $osp(1,2)$, 
is isomorphic to $sp(2, R)$. From (10) it is obvious that, if and only 
if the action $S_m$ is $Sp(2)$-invariant, it can be (anti)BRST-invariant as 
well. For the generators $\sigma_\alpha$ we may choose the representation 
\begin{equation}
\sigma_0 = \tau_3,
\qquad
\sigma_\pm = - \tau_\pm = - \hbox{$\frac{1}{2}$} (\tau_1 \pm i \tau_2),
\end{equation}
where $\tau_\alpha$ ($\alpha = 1,2,3$) are the Pauli matrices.

In order that the (anti)commutation relations (8)--(10) are consistent the 
structure constants $\epsilon_{\alpha\beta}^{~~~\!\gamma}$, 
$(\sigma_\alpha)_b^{~a}$ and $(\sigma^\alpha)^{ab}$ on the right-hand side
of these relations are subject to certain conditions which follow from the 
(graded) Jacobi identities. The requirement that the full set of these 
identities, namely
\begin{align*}
[ \bar{\Delta}_\alpha, [ \bar{\Delta}_\beta,\bar{\Delta}_\gamma ] ] +
[ \bar{\Delta}_\beta, [ \bar{\Delta}_\gamma,\bar{\Delta}_\alpha ] ] +
[ \bar{\Delta}_\gamma, [ \bar{\Delta}_\alpha,\bar{\Delta}_\beta ] ] &= 0,
\\
[ \bar{\Delta}_\alpha, [ \bar{\Delta}_\beta,\bar{\Delta}_m^a ] ] -
[ \bar{\Delta}_\beta, [ \bar{\Delta}_\alpha,\bar{\Delta}_m^a ] ] -
[ [ \bar{\Delta}_\alpha,\bar{\Delta}_\beta], \bar{\Delta}_m^a ] &= 0,
\\
\{ \bar{\Delta}_m^a, [ \bar{\Delta}_\alpha,\bar{\Delta}_m^b ] \} +
\{ \bar{\Delta}_m^b, [ \bar{\Delta}_\alpha,\bar{\Delta}_m^a ] \} +
[ \{ \bar{\Delta}_m^a,\bar{\Delta}_m^b \}, \bar{\Delta}_\alpha ] &= 0, 
\\
[ \bar{\Delta}_m^a, \{ \bar{\Delta}_m^b,\bar{\Delta}_m^c \} ] +
[ \bar{\Delta}_m^b, \{ \bar{\Delta}_m^c,\bar{\Delta}_m^a \} ] +
[ \bar{\Delta}_m^c, \{ \bar{\Delta}_m^a,\bar{\Delta}_m^b \} ] &= 0
\end{align*}
be fulfilled is equivalent to the demand that in the adjoint representation of 
$\bar{\Delta}_\alpha$ and $\bar{\Delta}_m^a$,
\begin{equation*}
\bar{\Delta}_\alpha = \begin{pmatrix} R_\alpha & 0 \\
                                      0 & S_\alpha \end{pmatrix},
\qquad
\bar{\Delta}_m^a = m \begin{pmatrix} 0 & U^a \\
                                     T^a & 0 \end{pmatrix}, 
\end{equation*}
the matrices $R_\alpha$ and $S_\alpha$ should form a representation of the 
(even) subalgebra $sl(2)$ of $osp(1,2)$. The elements of these matrices are 
\begin{equation*}
(R_{\alpha})_\beta^{~\gamma} = \epsilon_{\beta\alpha}^{~~~\!\gamma},
\qquad
(S_{\alpha})_a^{~b} = (\sigma_{\alpha})_a^{~b},
\qquad
(T^a)_b^{~\alpha} = i (\sigma^{\alpha})_b^{~a},
\qquad
(U^a)_\alpha^{~b} = - i (\sigma_{\alpha})^{ab}.
\end{equation*}
Therefore, we have the following conditions: (I) the matrices $R_\alpha$
form the adjoint representation of $sl(2)$ and (II) the 
matrices $S_\alpha$ form the fundamental representation of this algebra; 
(III) the matrix elements $(T^a)_b^{~\alpha}$ and $(U^a)_\alpha^{~b}$ are 
numerical invariants under $sl(2)$,
\begin{equation*}
(T^a)_b^{~\gamma} \epsilon_{\gamma\alpha}^{~~~\!\beta} =
(\sigma_\alpha)_b^{~c} (T^a)_c^{~\beta} -
(T^c)_b^{~\beta} (\sigma_\alpha)_c^{~a},
\qquad
\epsilon_{\alpha\beta}^{~~~\!\gamma} (U^a)_\gamma^{~b} =
(\sigma_\alpha)^a_{~c} (U^c)_\beta^{~b} -
(U^a)_\beta^{~c} (\sigma_\alpha)_c^{~b};
\end{equation*}
and (IV) there is a cyclic identity, 
\begin{equation*}
(\sigma^\alpha)^{ab} (\sigma_\alpha)^c_{~d} + 
\hbox{cyclic perm}(a, b, c) = 0,
\end{equation*}
which restricts the possible representations $\bar{\Delta}_m^a$ of the 
algebra $sl(2)$ spanned by $\bar{\Delta}_\alpha$ (in general, not every 
ordinary Lie algebra can be extended to a superalgebra). 

Let us notice that by invoking $osp(1,2)$-symmetry the notion of ghost number 
will be a natural property of the superalgebra (8)--(10). Indeed, from (8) 
and (9) we observe the relations 
\begin{equation*}
(\hbar/ i) [ \bar{\Delta}_0, \bar{\Delta}_\pm ] = 
\pm 2 \bar{\Delta}_\pm,
\qquad
(\hbar/ i) [ \bar{\Delta}_0, \bar{\Delta}^a_m ] =
\bar{\Delta}^b_m (\sigma_0)_b^{~a},
\end{equation*}
showing that $(\hbar/ i) \bar{\Delta}_0 = \Delta_{gh}$ is the 
Faddeev-Popov operator whose eigenvalues $gh(X)$ define the ghost numbers 
of the corresponding quantities $X$ of the theory in question,\break i.e. 
$\bigr([ \Delta_{gh}, X ] - gh(X) X \bigr)\, {\rm exp}\{(i/ \hbar) S_m\} = 0$.
 
However, insisting on $osp(1,2)$-symmetry this approach brings in a 
fundamentally new aspect. Namely, in order to express the superalgebra 
(8)--(10) by {\it operator identities} and to get nontrivial solutions of the
generating equations (6) and (7) one is enforced to enlarge the set of
antifields. More precisely, because the (anti)ghost fields $C^{\alpha_0 a_0}$ 
as well as their related antifields transform under $Sp(2)$ in a nontrivial 
way one has to introduce additional sources,
\begin{equation*}
\eta_A = (D_i, E_{\alpha_0}, F_{\alpha_0 a_0}),
\qquad
\epsilon(\eta_A) = \epsilon_A.
\end{equation*}
Let us remark that $D_i$ always could be set equal to zero (here also 
$E_{\alpha_0}$ has been intro-\break 
duced only for the sake of completeness and formal analogy to other fields -- 
it could be chosen equal to zero, too). For irreducible theories 
$F_{\alpha_0 a_0}$ is necessary in order to close the extended BRST algebra.
\footnote{For $L$-stage reducible theories both sources 
$F_{\alpha_s|a_0 \cdots a_s}$ and $E_{\alpha_s|a_1 \cdots a_s}$ 
($s = 0, \ldots , L$) are necessary in order to close the extended BRST 
algebra with respect to the more general space of auxiliary and (anti)ghost 
fields $B_{\alpha_s a| a_1 \cdots a_s}^*$, $C_{\alpha_s a| a_0 \cdots a_s}^*$,
${\bar B}_{\alpha_s| a_1 \cdots a_s}$, ${\bar C}_{\alpha_s| a_0 \cdots a_s}$.}

In order to set up the gauge fixing the new generalized gauge fixed quantum 
action $S_{m, {\rm ext}} = S_{m, {\rm ext}}(
\phi^A, \phi^*_{A a}, \bar{\phi}_A, \eta_A)$ will be introduced according to 
\begin{equation}
{\rm exp}\{ (i/ \hbar) S_{m, {\rm ext}} \} = 
\hat{U}_m(F) \,{\rm exp}\{ (i/ \hbar) S_m \}, 
\end{equation}
where the operator $\hat{U}_m(F)$ has to be choosen as 
\begin{equation}
\hat{U}_m(F) = {\rm exp}\{ (\hbar/ i) \hat{T}_m(F) \},
\qquad
\hat{T}_m(F) = \hbox{$\frac{1}{2}$} \epsilon_{ab} \{ 
\bar{\Delta}_m^b, [ \bar{\Delta}_m^a, F ] \} + (i/ \hbar)^2 m^2 F,
\end{equation}
$F = F(\phi^A)$ being the gauge fixing functional. Then, by virtue of (9) 
and (10), one establishes the relations
\begin{equation}
[ \bar{\Delta}_m^a, \hat{T}_m(F) ] = \hbox{$\frac{1}{2}$} 
(i/ \hbar) m^2 (\sigma^\alpha)^a_{~b} 
[ \bar{\Delta}_m^b, [ \bar{\Delta}_\alpha, F ] ]
\end{equation}
and
\begin{equation}
[ \bar{\Delta}_\alpha, \hat{T}_m(F) ] = \hbox{$\frac{1}{2}$} \epsilon_{ab} 
\{ \bar{\Delta}_m^b, [ 
\bar{\Delta}_m^a, [ \bar{\Delta}_\alpha, F ] ] \} +
(i/ \hbar)^2 m^2 [ \bar{\Delta}_\alpha, F ].
\end{equation}
Restricting $F$ to be a $Sp(2)$ scalar by imposing  
\begin{equation}
[ \bar{\Delta}_\alpha, F ] S_m = 0,
\end{equation}
it can be verified by using the explicit expressions of 
$\bar{\Delta}^a$ and $\bar{\Delta}_\alpha$ that the commutators
$[ \bar{\Delta}_m^a, \hat{U}_m(F) ] {\rm exp}\{ (i/ \hbar) S_m \} = 0$ and
$[ \bar{\Delta}_\alpha, \hat{U}_m(F) ] {\rm exp}\{ (i/ \hbar) S_m \} = 0$
vanish on the subspace of admissible actions $S_m$ (the proof of this 
statement will be postponed to section III).
Hence, the gauge fixed action $S_{m, {\rm ext}}$ satisfies eqs. (6) and (7) 
as well,
\begin{equation}
\bar{\Delta}_m^a\, {\rm exp}\{ (i/ \hbar) S_{m, {\rm ext}} \} = 0,
\qquad
\bar{\Delta}_\alpha\, {\rm exp}\{ (i/ \hbar) S_{m, {\rm ext}} \} = 0.
\end{equation}
Here, some remarks are in order. The operator $\bar{\Delta}_m^a$ is a 
nonlinear one; to be able to express the superalgebra (8)--(10) through 
operator identities the operator $\bar{\Delta}_\alpha$ must be a nonlinear 
operator, too. Therefore, it is not possible to impose the strong operator 
equation $[ \bar{\Delta}_\alpha, F ] = 0$, rather it must be replaced by the 
weaker condition $[ \bar{\Delta}_\alpha, F ] S_m = 0$. As a consequence, in 
order to prove that $S_{m, {\rm ext}}$ possesses the same symmetry properties 
as $S_m$ one needs the explicit realization of the operators  
$\bar{\Delta}_m^a$ and $\bar{\Delta}_\alpha$.

In this way the $Sp(2)$-covariant approach is generalized to another one based 
on the superalgebra $osp(1,2)$. Moreover, in this approach one can introduce 
a mass $m$ (which is necessary at least intermediately in the process of 
BPHZL-renormalization) without breaking the extended BRST symmetry.
\pagebreak
\bigskip\medskip
\begin{flushleft}
{\large{\bf III. EXPLICIT CONSTRUCTION OF THE
\\
$\phantom{\bf III.}$ GENERATING OPERATORS OF osp(1,2)}}
\end{flushleft}
\bigskip
After having stated the general structure of the $osp(1,2)$ quantization 
procedure being an obvious extension of $Sp(2)$ quantization we have to find 
an operational realization of the general operators just introduced. 
Thereby we proceed in such away that all formulars hold also for reducuible
gauge theories, except for the definition of the operator $(P_+)^{B a}_{A b}$
introduced below which has to be generalized. The explicit expressions for 
the operators $\bar{\Delta}_m^a$ and $\bar{\Delta}_\alpha$ in the generating 
equations (6) and (7) will be determined in two steps: First we construct a 
functional $S_m$ (at lowest order of $\hbar$) which is linear with respect to 
the antifields and is invariant under both (anti)BRST- and $Sp(2)$ 
transformations; later on we generalize to the case of nonlinear dependence. 
The corresponding (linear) symmetry operators being denoted by 
$\mathbf{s}_m^a$ and $\mathbf{d}_\alpha$ are required to fulfil the 
$osp(1,2)$-superalgebra:
\begin{equation*}
[ \mathbf{d}_\alpha, \mathbf{d}_\beta ] =  
\epsilon_{\alpha\beta}^{~~~\!\gamma} \mathbf{d}_\gamma,
\qquad
[ \mathbf{d}_\alpha, \mathbf{s}_m^a ] = 
\mathbf{s}_m^b (\sigma_\alpha)_b^{~a},
\qquad
\{ \mathbf{s}_m^a, \mathbf{s}_m^b \} = - m^2 (\sigma^\alpha)^{ab} 
\mathbf{d}_\alpha.
\end{equation*}
Let us make for $S_m$ the following ansatz:
\begin{equation}
S_m = S_{\rm cl} + W_X,
\qquad
W_X = (\hbox{$\frac{1}{2}$} \epsilon_{ab} 
\mathbf{s}_m^b \mathbf{s}_m^a + m^2) X, 
\end{equation}
where $X$ is assumed to be the following $Sp(2)$-scalar (in fact the only 
one we are able to build up linear in the antifields), 
\begin{equation*}
X = \bar{\phi}_A \phi^A,
\qquad 
\mathbf{d}_\alpha X = 0.
\end{equation*}
Then, by virtue of 
\begin{equation*}
\mathbf{s}_m^c ( \hbox{$\frac{1}{2}$} \epsilon_{ab} 
\mathbf{s}_m^b \mathbf{s}_m^a + m^2) X = \hbox{$\frac{1}{2}$} m^2
(\sigma^\alpha)^c_{~d} \mathbf{s}_m^d \mathbf{d}_\alpha X,
\qquad
[ \mathbf{d}_\alpha, \hbox{$\frac{1}{2}$} \epsilon_{ab} 
\mathbf{s}_m^b \mathbf{s}_m^a + m^2 ] X = 0,
\end{equation*}
it follows that $S_m$ is both (anti)BRST- and 
$Sp(2)$-invariant, $\mathbf{s}_m^a S_m = 0$ and $\mathbf{d}_\alpha S_m = 0$. 
Thereby, it is taken into account that due to gauge invariance (1) of 
$S_{\rm cl}(A)$ it holds $\mathbf{s}_m^a S_{\rm cl}(A) = 0$ with 
$\mathbf{s}_m^a A^i = R_{\alpha_0}^i C^{{\alpha_0} a}$ (here, the action of 
$\mathbf{s}_m^a$ -- and also of $\mathbf{d}_\alpha$ -- on the auxiliary 
and (anti)ghost fields is irrelevant and already left open).

The strategy to define the operators 
$\bar{\Delta}_m^a = \Delta^a + (i/ \hbar) V_m^a$, 
$\bar{\Delta}_\alpha = \Delta_\alpha + (i/ \hbar) V_\alpha$ is governed by a 
specific realization of the (anti)BRST- and $Sp(2)$-transformations of the 
{\it antifields} (extending the standard definitions of \cite{1}). 
Therefore, let us decompose $\mathbf{s}_m^a$ and $\mathbf{d}_\alpha$ into a 
component acting on the fields and another one acting on the antifields as 
follows:
\begin{equation*}
\mathbf{s}_m^a = 
(\mathbf{s}_m^a \phi^A) \frac{\delta_L}{\delta \phi^A} + V_m^a,
\qquad
\mathbf{d}_\alpha = 
(\mathbf{d}_\alpha \phi^A) \frac{\delta_L}{\delta \phi^A} + V_\alpha.
\end{equation*}
The action of $V_m^a$ and $V_\alpha$ on $\bar{\phi}_A$, $\phi^*_{A a}$ and 
$\eta_A$ are uniquely defined by 
\begin{alignat}{2}
V_m^a \bar{\phi}_A &= \epsilon^{ab} \phi^*_{A b},
& \qquad
V_\alpha \bar{\phi}_A &= 
\bar{\phi}_B (\sigma_\alpha)^B_{~A},
\nonumber\\
V_m^a \phi_{A b}^* &= m^2 (P_+)^{B a}_{A b} \bar{\phi}_B - 
\delta^a_b \eta_A, 
& \qquad
V_\alpha \phi^*_{A a} &=
\phi^*_{A b} (\sigma_\alpha)^b_{~a} + \phi^*_{B a} (\sigma_\alpha)^B_{~A},
\\
V_m^a \eta_A &= - m^2 \epsilon^{ab} (P_-)^{B c}_{A b} \phi^*_{B c},
& \qquad
V_\alpha \eta_A &= \eta_B (\sigma_\alpha)^B_{~A},
\nonumber
\end{alignat}
where the following abbreviations will be used:
\begin{equation*}
(P_-)^{B a}_{A b} \equiv (P_+)^{B a}_{A b} - 
(P_+)^B_A \delta^a_b + \delta^B_A \delta^a_b,
\qquad
(P_+)^B_A \equiv \delta^b_a (P_+)^{B a}_{A b},
\qquad
(\sigma_\alpha)^B_{~A} \equiv (\sigma_\alpha)^b_{~a} (P_+)^{B a}_{A b}.
\end{equation*}
Here, we introduced the matrix 
\begin{equation*}
(P_+)^{B a}_{A b} \equiv \begin{cases}
\delta^i_j \delta^a_b
& \text{for $A = i, B = j$},
\\ 
\delta_{\alpha_0}^{\beta_0} \delta^a_b
& \text{for $A = \alpha_0, B = \beta_0$},
\\
\delta_{\alpha_0}^{\beta_0} 2 (S_+)^{b_0 a}_{a_0 b}
& \text{for $A = \alpha_0 a_0, B = \beta_0 b_0$},
\\
0 & \text{otherwise,} 
\end{cases}
\end{equation*}
where the symmetrizator $(S_+)^{b_0 a}_{a_0 b}$ is defined as
\begin{equation*}
(S_+)^{b_0 a}_{a_0 b} \equiv \hbox{$\frac{1}{2}$}
\frac{\partial}{\partial X^{a_0}} \frac{\partial}{\partial X^b} X^a X^{b_0} = 
\hbox{$\frac{1}{2}$} (
\delta^a_b \delta^{a_0}_{b_0} + \delta^a_{a_0} \delta^{b_0}_b ),
\end{equation*}
$X^a$ being independent bosonic variables.

The matrices $(P_-)^B_A \equiv \delta^a_b (P_-)^{B a}_{A b}$ and 
$(\sigma_\alpha)^B_{~A}$ act nontrivially on the components of the 
(anti)fields having an internal (dummy) $Sp(2)$ index. For example, 
\begin{equation*}
(P_-)^B_A \bar{\phi}_B = 
(0, 0, - \bar{C}_{\alpha_0 a_0}),
\qquad
\bar{\phi}_B (\sigma_\alpha)^B_{~A} =
(0, 0, \bar{C}_{\alpha_0 c} (\sigma_\alpha)^c_{~a_0}).
\end{equation*}
Therefore, $V_\alpha$ acts only on the (anti)ghost-antifields, and $V_m^a$
is partly of that kind. Of course, we could have used also a componentwise 
notation, however, then the equations would be less easy to survey 
(a componentwise notation of the transformations (19) can be found 
in Appendix E).

In order to prove that the transformations (19) obey the 
$osp(1,2)$-superalgebra 
\begin{equation}
[ V_\alpha, V_\beta ] = \epsilon_{\alpha\beta}^{~~~\!\gamma} V_\gamma,
\qquad
[ V_\alpha, V_m^a ] = V_m^b (\sigma_\alpha)_b^{~a},
\qquad
\{ V_m^a, V_m^b \} = - m^2 (\sigma^\alpha)^{ab} V_\alpha
\end{equation}
one needs the following two equalities:
\begin{align}
\epsilon^{ad} (P_+)^{B b}_{A d} + \epsilon^{bd} (P_+)^{B a}_{A d} &=
- (\sigma^\alpha)^{ab} (\sigma_\alpha)^B_{~A},
\nonumber\\
\epsilon^{ad} (P_+)^{B b}_{A c} + \epsilon^{bd} (P_+)^{B a}_{A c} -
(\sigma^\alpha)^{ab} (\sigma_\alpha)^e_{~c} (P_-)^{B d}_{A e} &= 
- (\sigma^\alpha)^{ab} \bigr(
(\sigma_\alpha)^d_{~c} \delta^B_A + \delta^d_c (\sigma_\alpha)^B_{~A} \bigr),
\nonumber\\
\intertext{or equivalently,}
\epsilon^{ad} \delta^b_c + \epsilon^{bd} \delta^a_c &=
- (\sigma^\alpha)^{ab} (\sigma_\alpha)^d_{~c},
\\
(\sigma^\alpha)^{ab} \bigr(
(\sigma_\alpha)^{b_0}_{~~\!c} \delta^d_{a_0} + 
\delta^{b_0}_c (\sigma_\alpha)^d_{~a_0} \bigr) &=
(\sigma^\alpha)^{ab} \bigr(
(\sigma_\alpha)^d_{~c} \delta^{b_0}_{a_0} + 
\delta^d_c (\sigma_\alpha)^{b_0}_{~~a_0} \bigr), 
\end{align}
and the relation $(P_-)^{A b}_{C d} (P_+)^{C d}_{B a} = 0$. 

Then from (18) one gets for $S_m$ the expression
\begin{equation}
S_m = S_{\rm cl} + 
(\eta_A - \hbox{$\frac{1}{2}$} m^2 (P_+)^B_A \bar{\phi}_B) \phi^A -
(\mathbf{s}_m^a \phi^A) \phi^*_{A a} + \bar{\phi}_A ( 
\hbox{$\frac{1}{2}$} \epsilon_{ab} \mathbf{s}_m^b \mathbf{s}_m^a + m^2 )
\phi^A.
\end{equation}

Now, the symmetry properties of $S_m$ with respect to (anti)BRST- and 
$Sp(2)$-trans-\break
formations may be expressed by the following set of equations: 
\begin{equation}
\hbox{$\frac{1}{2}$} (S_m, S_m)^a + V_m^a S_m = 0,
\qquad
\hbox{$\frac{1}{2}$} \{ S_m, S_m \}_\alpha + V_\alpha S_m = 0;
\end{equation}
here the extended antibrackets $( F,G )^a$, introduced for the first time
in Ref. \cite{1}, define an odd graded and $\{ F,G \}_\alpha$ a new even 
graded algebraic structure on the space of fields and antifields 
(see Appendix B):
\begin{equation}
( F,G )^a = 
\frac{\delta F}{\delta \phi^A}
\frac{\delta G}{\delta \phi^*_{A a}} - 
(-1)^{(\epsilon(F) + 1)(\epsilon(G) + 1)} 
\frac{\delta G}{\delta \phi^A}
\frac{\delta F}{\delta \phi^*_{A a}}
\end{equation}
and
\begin{equation}
\{ F,G \}_\alpha = (\sigma_\alpha)_B^{~~\!A} \bigr(  
\frac{\delta F}{\delta \phi^A} \frac{\delta G}{\delta \eta_B} + 
(-1)^{\epsilon(F) \epsilon(G)}   
\frac{\delta G}{\delta \phi^A} \frac{\delta F}{\delta \eta_B} \bigr),
\end{equation}
and, in accordance with (19), the first-order differential operators 
$V_m^a$ and $V_\alpha$ are given by
\begin{equation}
V_m^a = 
\epsilon^{ab} \phi^*_{A b} \frac{\delta}{\delta \bar{\phi}_A} -
\eta_A \frac{\delta}{\delta \phi^*_{A a}} +
m^2 (P_+)^{B a}_{A b} \bar{\phi}_B \frac{\delta}{\delta \phi^*_{A b}} -
m^2 \epsilon^{ab} (P_-)^{B c}_{A b} \phi^*_{B c} \frac{\delta}{\delta \eta_A}
\end{equation}
and
\begin{equation}
V_\alpha =   
\bar{\phi}_B (\sigma_\alpha)^B_{~A} 
\frac{\delta}{\delta \bar{\phi}_A} + \bigr( 
\phi^*_{A b} (\sigma_\alpha)^b_{~a} +
\phi^*_{B a} (\sigma_\alpha)^B_{~A} \bigr)
\frac{\delta}{\delta \phi^*_{A a}} + 
\eta_B (\sigma_\alpha)^B_{~A} 
\frac{\delta}{\delta \eta_A}. 
\end{equation}
We also introduce the second-order differential operators $\Delta^a$ 
and $\Delta_\alpha$ whose structure is extracted from (25) and (26):
\begin{equation}
\Delta^a = (-1)^{\epsilon_A}  
\frac{\delta_L}{\delta \phi^A}
\frac{\delta}{\delta \phi^*_{A a}}, 
\qquad
\Delta_\alpha = (-1)^{\epsilon_A} (\sigma_\alpha)_B^{~~\!A} 
\frac{\delta_L}{\delta \phi^A}
\frac{\delta}{\delta \eta_B}.
\end{equation}

Let us now consider the general case where the action $S_m$ is assumed to 
appear in\break the form of a series expansion in powers of $\hbar$ which 
may depend also {\it nonlinear} on the antifields. This action will be 
required to satisfy the following set of quantum master equations
(see Appendix C):
\begin{equation}
\hbox{$\frac{1}{2}$} (S_m, S_m)^a + V_m^a S_m =
i \hbar \Delta^a S_m,
\qquad
\hbox{$\frac{1}{2}$} \{ S_m, S_m \}_\alpha + V_\alpha S_m =
i \hbar \Delta_\alpha S_m,
\end{equation}
or equivalently,
\begin{equation}
\bar{\Delta}_m^a\, {\rm exp}\{ (i/ \hbar) S_m \} = 0,
\qquad
\bar{\Delta}_\alpha\, {\rm exp}\{ (i/ \hbar) S_m \} = 0.
\end{equation}
Furthermore, by an explicit calculation it can be verified that  
$\bar{\Delta}_m^a = \Delta^a + (i/ \hbar) V_m^a$ and
$\bar{\Delta}_\alpha = \Delta_\alpha + (i/ \hbar) V_\alpha$ obey the 
$osp(1,2)$-superalgebra (8)--(10).

In order to lift the degeneracy of $S_m$ we follow the general gauge-fixing 
procedure suggested by (12) and (13). Let us introduce the gauge fixed action
\begin{equation}
{\rm exp}\{ (i/ \hbar) S_{m, {\rm ext}} \} = 
\hat{U}_m(F)\, {\rm exp}\{ (i/ \hbar) S_m \},
\end{equation}
where the operator $\hat{U}_m(F) = {\rm exp}\{ (i/ \hbar) \hat{T}_m(F) \}$ 
is defined according to the formula (13). If the gauge-fixing functional 
is assumed to depend only on the fields, $F = F(\phi^A)$, then one gets
\begin{equation}
\hat{U}_m(F) = {\rm exp}\bigr\{
\frac{\delta F}{\delta \phi^A} (
\frac{\delta}{\delta \bar{\phi}_A} -
\hbox{$\frac{1}{2}$} m^2 (P_-)^A_B \frac{\delta}{\delta \eta_B} ) -
(\hbar/ i) \hbox{$\frac{1}{2}$} \epsilon_{ab} 
\frac{\delta}{\delta \phi^*_{A a}}
\frac{\delta^2 F}{\delta \phi^A \delta \phi^B}
\frac{\delta}{\delta \phi^*_{B b}} + (i/ \hbar) m^2 F \bigr\}.
\end{equation}
Let us prove that $S_{m, {\rm ext}}$ obeys the generating equations 
(30) and (31) as well. Clearly, since $\bar{\Delta}_m^a$, 
$\bar{\Delta}_\alpha$ and $\hat{U}_m(F)$ do not commute with each other
this proof will be more involved than in the BLT approach. This is due to the 
fact that, looking back at (14) and (15), neither
\begin{equation*}
[ \bar{\Delta}_m^a, \hat{T}_m(F) ] = \hbox{$\frac{1}{2}$} 
(i/ \hbar) m^2 (\sigma_\alpha)^a_{~b} 
[ \bar{\Delta}_m^b, [ \bar{\Delta}^\alpha, F ] ]
\end{equation*}
nor
\begin{equation*}
[ \bar{\Delta}_\alpha, \hat{T}_m(F) ] = \hbox{$\frac{1}{2}$} \epsilon_{ab} 
\{ \bar{\Delta}_m^b, [ 
\bar{\Delta}_m^a, [ \bar{\Delta}_\alpha, F ] ] \} +
(i/ \hbar)^2 m^2 [ \bar{\Delta}_\alpha, F ]
\end{equation*}
does vanish, since due to the nonlinearity of $\bar{\Delta}_\alpha$ one cannot 
require the strong condition $[ \bar{\Delta}_\alpha, F ] = 0$.  However, 
a direct verification shows that $\hat{T}_m(F)$ commutes with any term on 
the right-hand side of both previous relations, i.e. it holds
\begin{equation}
[ \hat{T}_m(F), [ \bar{\Delta}_m^a, \hat{T}_m(F) ] ] = 0,
\qquad
[ \hat{T}_m(F), [ \bar{\Delta}_\alpha, \hat{T}_m(F) ] ] = 0.
\end{equation}
Then, by the help of (34) one obtains
\begin{equation*}
[ \bar{\Delta}_m^a, \hat{U}_m(F) ] = 
(\hbar/ i) \hat{U}_m(F) [ \bar{\Delta}_m^a, \hat{T}_m(F) ],
\qquad
[ \bar{\Delta}_\alpha, \hat{U}_m(F) ] = 
(\hbar/ i) \hat{U}_m(F) [ \bar{\Delta}_\alpha, \hat{T}_m(F) ].
\end{equation*}
Let us require
\begin{equation*}
[ \bar{\Delta}_\alpha, F ] S_m \equiv (\sigma_\alpha)_B^{~~\!A}
\frac{\delta F}{\delta \phi^A} \frac{\delta}{\delta \eta_B} S_m = 0,
\end{equation*}
then, taking into account that $S_m$ solves the generating equations (31), 
it is easily seen that $[ \bar{\Delta}_m^a, \hat{U}_m(F) ]$ and 
$[ \bar{\Delta}_\alpha, \hat{U}_m(F) ]$ vanish after acting on 
${\rm exp}\{ (i/ \hbar) S_m \}$,
\begin{equation*} 
[ \bar{\Delta}_m^a, \hat{U}_m(F) ]\, {\rm exp}\{ (i/ \hbar) S_m \} = 0,
\qquad 
[ \bar{\Delta}_\alpha, \hat{U}_m(F) ]\, {\rm exp}\{ (i/ \hbar) S_m \} = 0.
\end{equation*}
Summarizing, we have the results
\begin{equation}
\bar{\Delta}_m^a\, {\rm exp}\{ (i/ \hbar) S_{m, {\rm ext}} \} = 0,
\qquad
\bar{\Delta}_\alpha\, {\rm exp}\{ (i/ \hbar) S_{m, {\rm ext}} \} = 0,
\end{equation}
i.e. the gauge-fixed action $S_{m, {\rm ext}}$ satisfies the same generating 
equations (30) and (31) as $S_m$ which is indeed what we intended to prove.
\bigskip\medskip
\begin{flushleft}
{\large{\bf IV. WARD IDENTITIES, GENERATING FUNCTIONALS 
\\
$\phantom{\bf IV.}$ AND GAUGE (IN)DEPENDENCE}} 
\end{flushleft}
\bigskip
Next, we turn to the question of gauge independence of physical quantities, 
especially of the $S$-matrix. In discussing this question it is convenient to 
study first the symmetry properties of the vacuum functional, 
\begin{equation}
Z_m(0) = \int d \phi^A \,
{\exp}\{ (i/ \hbar) S_{m, {\rm eff}}(\phi^A) \},
\end{equation} 
where $S_{m, {\rm eff}}(\phi^A) = 
S_{m, {\rm ext}}(\phi^A, \phi_{A a}^*, \bar{\phi}_A, \eta_A)
|_{\phi^*_a = \bar{\phi} = \eta = 0}$. It can be represented in the form
\begin{equation}
Z_m (0)= \int d \phi^A\, d \eta_A\, d \phi^*_{A a}\, 
d \pi^{A a}\, d \bar{\phi}_A\, d \lambda^A\, \delta(\eta_A)\, 
{\exp}\{ (i/ \hbar) (S_{m, {\rm ext}} - W_X) \}
\end{equation} 
with
\begin{equation}
W_X = (\eta_A - \hbox{$\frac{1}{2}$} m^2 (P_+)^B_A \bar{\phi}_B) \phi^A -
\phi^*_{A a} \pi^{A a} -
\bar{\phi}_A (\lambda^A - \hbox{$\frac{1}{2}$} m^2 (P_-)^A_B \phi^B).
\end{equation}
where we have extended the space of variables by introducing the auxiliary 
fields $\pi^{A a}$ and $\lambda^A$. It is straightforward to check that $W_X$ 
can be cast into the form (38). Indeed, performing in (37) first of all
the integrations over $\lambda^A$ and $\pi^{A a}$, which yields the
delta-functions $\delta(\bar{\phi}_A)$ and $\delta(\phi_{A a}^*)$, and,
after that, carring out the integrations over $\bar{\phi}_A$, $\phi_{A a}^*$ 
and $\eta_A$ one gets the expression (36).      

Since $\eta_A$ for $m \neq 0$ transforms nontrivially under $osp(1,2)$ 
we express $\delta(\eta_A)$ by 
\begin{equation*}
\delta(\eta_A) = \int d \zeta^A\, {\exp}\{ (i/ \hbar) \eta_A \zeta^A \} 
\end{equation*}
and change within (37) the integration variables $\phi^A$ and $\lambda^A$ 
according to $\phi^A \rightarrow \phi^A + \zeta^A$ and 
$\lambda^A \rightarrow \lambda^A + \hbox{$\frac{1}{2}$} m^2 (
(P_-)^A_B - (P_+)^A_B) \zeta^B$. Then, for $Z_m(0)$ this yields
\begin{equation}
Z_m (0) = \int d \phi^A\, d \eta_A\, d \zeta^A\, 
d \phi^*_{A a}\, d \pi^{A a}\, d \bar{\phi}_A\, d \lambda^A\, 
{\exp}\{ (i/ \hbar) (S_{m, {\rm ext}}^\zeta - W_X) \},
\end{equation}
where $S_{m, {\rm ext}}^\zeta$ is obtained from $S_{m, {\rm ext}}$ by
performing the replacement $\phi^A \rightarrow \phi^A + \zeta^A$.
At this stage we remark that ($\phi^A$, $\pi^{A a}$, $\lambda^A$) and 
($\bar{\phi}_A$, $\phi^*_{A a}$, $\eta_A$) constitute the components of the 
superfield and superantifield, respectively, of the superfield quantization 
scheme \cite{9}; here we changed the notation 
$J_A \rightarrow \eta_A$ relative to \cite{9}. Of course, the formalism 
introduced here may be written in that form also.
 
The term $W_X$ may be cast into the $osp(1,2)$-invariant form
\begin{equation*}
W_X = ( \hbox{$\frac{1}{2}$} \epsilon_{ab} 
(V_m^b - U_m^b) (V_m^a - U_m^a) + m^2 ) X,
\qquad
X = \bar{\phi}_A \phi^A,
\qquad
(V_\alpha + U_\alpha) X = 0,
\end{equation*}
with $V_m^a$ and $V_\alpha$, whose action on $\bar{\phi}_A$, $\phi^*_{A a}$, 
$\eta_A$ are already defined in (19), satisfying the $osp(1,2)$-superalgebra
(20), and the action of $U_m^a$ and $U_\alpha$ on $\phi^A$, $\pi^{A a}$, 
$\lambda^A$ and $\zeta^A$ being defined according to
\begin{alignat}{2}
\phi^A U_m^a &= \pi^{A a},
& \qquad
\phi^A U_\alpha &= 
\phi^B (\sigma_\alpha)_B^{~~\!A},
\nonumber\\
\pi^{A b} U_m^a &= \epsilon^{ab} \lambda^A +
m^2 \epsilon^{ac} (P_+)^{A b}_{B c} \phi^B, 
& \qquad
\pi^{A a} U_\alpha &=
\pi^{A b} (\sigma_\alpha)_b^{~a} + \pi^{B a} (\sigma_\alpha)_B^{~~\!A},
\\
\lambda^A U_m^a &= m^2 (P_-)^{A a}_{B b} \pi^{B b},
& \qquad
\lambda^A U_\alpha &= \lambda^B (\sigma_\alpha)_B^{~~\!A},
\nonumber\\
\zeta^A U_m^a &= 0,
& \qquad
\zeta^A U_\alpha &= \zeta^B (\sigma_\alpha)_B^{~~\!A}.
\nonumber
\end{alignat}
Here, $U_m^a$ and $U_\alpha$ are defined as {\it right} derivatives, 
in contrast to $V_m^a$ and $V_\alpha$, which was defined as {\it left} ones. 
In order to prove that the transformations (40) obey the 
$osp(1,2)$-superalgebra 
\begin{equation}
[ U_\alpha, U_\beta ] = - \epsilon_{\alpha\beta}^{~~~\!\gamma} U_\gamma,
\qquad
[ U_\alpha, U_m^a ] = - U_m^b (\sigma_\alpha)_b^{~a},
\qquad
\{ U_m^a, U_m^b \} = m^2 (\sigma^\alpha)^{ab} U_\alpha
\end{equation}
one needs the following two equalities:
\begin{align*}
\epsilon^{ad} (P_+)^{B b}_{A d} + \epsilon^{bd} (P_+)^{B a}_{A d} &=
(\sigma^\alpha)^{ab} (\sigma_\alpha)_A^{~~\!B},
\\
\epsilon^{ad} (P_-)^{B b}_{A c} + \epsilon^{bd} (P_-)^{B a}_{A c} -
(\sigma^\alpha)^{ab} (\sigma_\alpha)^e_{~c} (P_+)^{B d}_{A e} &= 
(\sigma^\alpha)^{ab} \bigr(
(\sigma_\alpha)_c^{~d} \delta^B_A + 
\delta^d_c (\sigma_\alpha)_A^{~~\!B} \bigr),
\end{align*}
and the relation $(P_-)^{A b}_{C d} (P_+)^{C d}_{B a} = 0$. It is easily 
seen that both equalities are equivalent to (21) and (22), too.

Inserting into (39) the relations (32), (33) and integrating by parts this 
yields 
\begin{equation}
Z_m (0) = \int d \phi^A\, d \eta_A\, d \zeta^A\,
d \phi_{A a}^*\, d \pi^{A a}\, d \bar{\phi}_A\, d \lambda^A\, 
{\exp}\{ (i/ \hbar) (S_m^\zeta + W_F^\zeta - W_X) \}
\end{equation}
with the following expression for $W_F$:
\begin{equation}
W_F = - \frac{\delta F}{\delta \phi^A} (
\lambda^A + \hbox{$\frac{1}{2}$} m^2 (P_+)^A_B \phi^B ) - 
\hbox{$\frac{1}{2}$} \epsilon_{ab} 
\pi^{A a} \frac{\delta^2 F}{\delta \phi^A \delta \phi^B} \pi^{B b} + m^2 F
\end{equation}
which may be recast into the $osp(1,2)$-invariant form
\begin{equation*}
W_F = F ( \hbox{$\frac{1}{2}$} \epsilon_{ab} U_m^b U_m^a + m^2),
\qquad
F U_\alpha = 0.
\end{equation*}
(Again, $S_m^\zeta$ and $W_F^\zeta$ are obtained from $S_m$ and $W_F$,
respectively, by carrying out the replacement
$\phi^A \rightarrow \phi^A + \zeta^A$.)

Let us now introduce the combined (first-order) differential operators
\begin{equation*}
L_m^a X \equiv V_m^a X - X U_m^a,
\qquad
L_\alpha X \equiv V_\alpha X + X U_\alpha,
\end{equation*}
where $U_m^a$ and $U_\alpha$, in accordance with (40), are defined by
\begin{equation*}
U_m^a = 
\frac{\delta}{\delta \phi^A} 
\pi^{A a} + 
\frac{\delta}{\delta \pi^{A b}}
\epsilon^{ab} \lambda^A + 
\frac{\delta}{\delta \pi^{A b}}
m^2 \epsilon^{ac} (P_+)^{A b}_{B c} \phi^B +
\frac{\delta}{\delta \lambda^A}
m^2 (P_-)^{A a}_{B b} \pi^{B b} 
\end{equation*}
and
\begin{equation*}
U_\alpha = 
\frac{\delta}{\delta \phi^A}
\phi^B (\sigma_\alpha)_B^{~~\!A} +
\frac{\delta}{\delta \lambda^A}
\lambda^B (\sigma_\alpha)_B^{~~\!A} +
\frac{\delta}{\delta \pi^{A a}} \bigr(
\pi^{A b} (\sigma_\alpha)_b^{~a} + 
\pi^{B a} (\sigma_\alpha)_B^{~~\!A} \bigr) +
\frac{\delta}{\delta \zeta^A}
\zeta^B (\sigma_\alpha)_B^{~~\!A};
\end{equation*}
here, the derivatives with respect to $\phi^A$, $\pi^{A b}$, $\lambda^A$ and 
$\zeta^A$ are {\it right} ones ($V_m^a$ and $V_\alpha$ are already defined 
in (27) and (28)). Then, by virtue of (20) and (41), it follows
that $L_m^a$ and $L_\alpha$ satisfy the $osp(1,2)$-superalgebra
\begin{equation*}
[ L_\alpha, L_\beta ] = \epsilon_{\alpha\beta}^{~~~\!\gamma} L_\gamma,
\qquad
[ L_\alpha, L_m^a ] = L_m^b (\sigma_\alpha)_b^{~a},
\qquad
\{ L_m^a, L_m^b \} = - m^2 (\sigma^\alpha)^{ab} L_\alpha.
\end{equation*} 
Hence, it holds
\begin{equation}
L_m^c ( \hbox{$\frac{1}{2}$} \epsilon_{ab} L_m^b L_m^a + m^2 ) = 
\hbox{$\frac{1}{2}$} m^2 (\sigma^\alpha)^c_{~d} L_m^d L_\alpha,
\qquad
[ L_\alpha, \hbox{$\frac{1}{2}$} \epsilon_{ab} L_m^b L_m^a + m^2 ] = 0.
\end{equation}

We assert now that (42) is invariant under the following (global) 
transformations (thereby, one has to make use of the first equation in (30)
and (44), respectively):
\begin{alignat}{3}
\delta_m \phi^A &= \phi^A U_m^a \mu_a,
& \qquad
\delta_m \zeta^A &= 0,
& \qquad
\delta_m \bar{\phi}_A &= \mu_a V_m^a \bar{\phi}_A,
\nonumber\\
\delta_m \pi^{A b} &= \pi^{A b} U_m^a \mu_a,
& \qquad
&
& \qquad
\delta_m \phi^*_{A b} &= \mu_a V_m^a \phi^*_{A b} +
\mu_a ( S_m^\zeta, \phi^*_{A b} )^a,
\\
\delta_m \lambda^A &= \lambda^A U_m^a \mu_a,
& \qquad
&
& \qquad
\delta_m \eta_A &= \mu_a V_m^a \eta_A,
\nonumber
\end{alignat} 
where $\mu_a$, $\epsilon(\mu_a) = 1$, is a $Sp(2)$-doublet of constant 
anticommuting parameters. The transformations (45), formally generated by  
$\delta_m = \mu_a (V_m^a - U_m^a) + \mu_a ( S_m^\zeta, ~\cdot~ )^a$, realize 
the $m$-extended BRST symmetry in the space of variables 
$\phi^A$, $\bar{\phi}_A$, $\phi^*_{A a}$, $\eta_A$, $\pi^{A a}$, $\lambda^A$ 
and $\zeta^A$.

Moreover, it is straightforward to check that (42) is also invariant under the 
following transformations (where one has to make use of the second equation 
in (30) and (44), respectively):
\begin{alignat}{3}
\delta \phi^A &= \phi^A U_\alpha \theta^\alpha,
& \qquad
\delta \zeta^A &= \zeta^A U_\alpha \theta^\alpha,
& \qquad
\delta \bar{\phi}_A &= \theta^\alpha V_\alpha \bar{\phi}_A,
\nonumber\\
\delta \pi^{A b} &= \pi^{A b} U_\alpha \theta^\alpha,
& \qquad
&
& \qquad
\delta \phi^*_{A b} &= \theta^\alpha V_\alpha \phi^*_{A b},
\\
\delta \lambda^A &= \lambda^A U_\alpha \theta^\alpha,
& \qquad
&
& \qquad
\delta \eta_A &= \theta^\alpha V_\alpha \eta_A + 
\theta^\alpha \{ S_m^\zeta, \eta_A \}_\alpha,
\nonumber
\end{alignat}
where $\theta^\alpha$, $\epsilon(\theta^\alpha) = 0$, are constant commuting
parameters. The transformations (46), formally generated by $\delta = 
\theta^\alpha ( V_\alpha + U_\alpha ) + 
\theta^\alpha \{ S_m^\zeta, ~\cdot~ \}_\alpha$, realize the $Sp(2)$-symmetry 
in the space of variables $\phi^A$, $\bar{\phi}_A$, $\phi^*_{A a}$, $\eta_A$, 
$\pi^{A a}$, $\lambda^A$ and $\zeta^A$.

In principle, for a general gauge functional $F$, $\mu_a$ may be assumed to 
depend on all these variables $\phi^A$, $\bar{\phi}_A$, $\phi^*_{A a}$,
$\eta_A$, $\pi^{A a}$, $\lambda^A$ and $\zeta^A$. As long as $F$ depends 
only on the fields it is sufficient for $\mu_a$ to depend on $\phi^A$ and 
$\pi^{A a}$ only. Then the symmetry of the vacuum functional $Z_m(0)$ with 
respect to the transformations (45) permits to study the question 
whether the mass dependent terms of the action violate the independence of 
the $S$-matrix on the choice of the gauge. 

Indeed, let us change the gauge-fixing functional 
$F(\phi) \rightarrow F(\phi) + \delta F(\phi)$. Then the gauge-fixing term 
$W_F$ changes according to
\begin{equation}
W_F \rightarrow W_{F + \delta F} = W_F + W_{\delta F},
\qquad
W_{\delta F} = \delta F(\phi)
( \hbox{$\frac{1}{2}$} \epsilon_{ab} U_m^b U_m^a + m^2 ).
\end{equation}
Now, performing in (42) the transformations (45), we choose 
\begin{equation*}
\mu_a = \mu_a(\phi, \pi) \equiv
 - (i/ \hbar) \hbox{$\frac{1}{2}$} \epsilon_{ab} \delta F(\phi) U_m^b.
\end{equation*}
This induces the factor ${\rm exp}(\mu_a U_m^a)$ in the integration measure.
Combining its exponent with $W_F$ leads to
\begin{equation*}
W_F \rightarrow W_F + (\hbar/ i) \mu_a U_m^a =
W_F - \hbox{$\frac{1}{2}$} \epsilon_{ab} \delta F(\phi) U_m^b U_m^a = 
W_F - W_{\delta F} + m^2 \delta F(\phi).
\end{equation*}
By comparison with (47) we see that the mass term $m^2 F$ in $W_F$ violates 
the independence of the vacuum functional $Z_m(0)$ on the choice of the gauge. 
This result, together with the equivalence theorem \cite{10},
is sufficient to prove that the same is true also for the $S$-matrix. 

One may try to compensate this undesired term $m^2 \delta F(\phi)$ by means
of an additional change of variables. But this change should not
destroy the form of the action arrived at the previous stage. However,
an additional change of variables leads to a Berezinian which is equal to
one because $\sigma_\alpha$ are traceless. Therefore, the unwanted term
could never be compensated.

Finally, we shall derive the Ward identities for the extended BRST- and the 
$Sp(2)$-symmetries. To begin with, let us introduce the generating functional 
of the Green's functions:
\begin{equation}
Z_m(J_A; \phi^*_{A a}, \bar{\phi}_A, \eta_A) = \int d \phi^A\, 
{\exp}\bigr\{ (i/ \hbar) \bigr(
S_{m, {\rm ext}}(\phi^A, \phi^*_{A a}, \bar{\phi}_A, \eta_A) +
J_A \phi^A \bigr) \bigr\}.
\end{equation} 
If we multiply eqs. (35) from the left by 
${\rm exp}\{ (i/ \hbar) J_A \phi^A \}$ and integrate over $\phi^A$ we get
\begin{align*}
\int d \phi^A\, {\exp}\{ (i/ \hbar) J_A \phi^A \}
\bar{\Delta}_m^a\, {\exp}\bigr\{ (i/ \hbar) 
S_{m, {\rm ext}}(\phi^A, \phi^*_{A a}, \bar{\phi}_A, \eta_A) \bigr\} &= 0,
\\ 
\int d \phi^A\, {\exp}\{ (i/ \hbar) J_A \phi^A \}
\bar{\Delta}_\alpha\, {\exp}\bigr\{ (i/ \hbar) 
S_{m, {\rm ext}}(\phi^A, \phi^*_{A a}, \bar{\phi}_A, \eta_A) \bigr\} &= 0.
\end{align*}
Now, integrating by parts and assuming the integrated expressions to vanish, 
we can rewrite the resulting equalities by the help of the definition (48) as 
\begin{align*}
(J_A \frac{\delta}{\delta \phi^*_{A a}} - 
V_m^a) Z_m(J_A; \phi^*_{A a}, \bar{\phi}_A, \eta_A) &= 0,
\\
\bigr((\sigma_\alpha)_B^{~~\!A} J_A \frac{\delta}{\delta \eta_B} - 
V_\alpha\bigr) Z_m(J_A; \phi^*_{A a}, \bar{\phi}_A, \eta_A) &= 0,
\end{align*}
which are the Ward identities for the generating functional of Green's 
functions. 

Introducing as usual the generating functional of the vertex functions,
\begin{align*}
\Gamma_m(\phi^A; \phi^*_{A a}, \bar{\phi}_A, \eta_A) &=
(\hbar/ i)\, {\rm ln} Z_m(J_A; \phi^*_{A a}, \bar{\phi}_A, \eta_A) -
J_A \phi^A,
\\
\phi^A &= (\hbar/ i)
\frac{\delta {\rm ln} Z_m(J_A; \phi^*_{A a}, \bar{\phi}_A, \eta_A)}
{\delta J_A},
\end{align*}
we obtain
\begin{equation}
\hbox{$\frac{1}{2}$} ( \Gamma_m, \Gamma_m )^a + V_m^a \Gamma_m = 0,
\qquad
\hbox{$\frac{1}{2}$} \{ \Gamma_m, \Gamma_m \}_\alpha + V_\alpha \Gamma_m = 0.
\end{equation}
For Yang-Mills theories the first identities in (49) are the Slavnov-Taylor 
identities of the extended BRST symmetries. Furthermore, choosing for 
$\sigma_\alpha$ the representation (11) the second identities in (49) express 
for $\alpha = 0$ the ghost number conservation and, in Yang-Mills theories, 
for $\alpha = (+,-)$ the Delduc-Sorella identities of the $Sp(2)$-symmetry 
\cite{11}.
\bigskip\medskip
\begin{flushleft}
{\large{\bf V. MASSIVE THEORIES WITH CLOSED ALGEBRA}} 
\end{flushleft}
\bigskip
To illustrate the formalism of the $osp(1,2)$-quantization developed here, 
we consider irreducible massive gauge theories with a closed algebra. Such 
theories are characterized by the fact, first, that in the algebra of 
generators, eq. (2), it holds $M_{\alpha_0 \beta_0}^{ij} = 0$, i.e.
\begin{equation}
R^i_{\alpha_0, j} R^j_{\beta_0} - R^i_{\beta_0, j} R^j_{\alpha_0} = 
- R^i_{\gamma_0} F^{\gamma_0}_{\alpha_0 \beta_0},
\end{equation}
where for the sake of simplicity we assume throughout this and the succeding 
section that $A^i$ are {\it bosonic} fields. Second, that $R^i_{\alpha_0}$
has no zero modes (contrary to the case of reducible theories), i.e. any 
equation of the form $R^i_{\alpha_0} X^{\alpha_0} = 0$ has only the trivial 
solution $X^{\alpha_0} = 0$. Then in the case of field-dependent structure 
constants the Jacobi identity looks like
\begin{equation}
F^{\delta_0}_{\eta_0 \alpha_0} F^{\eta_0}_{\beta_0 \gamma_0} -
R^i_{\alpha_0} F^{\delta_0}_{\beta_0 \gamma_0, i} +
\hbox{cyclic perm}(\alpha_0, \beta_0, \gamma_0) = 0.
\end{equation}
We shall restrict ourselves to consider only solutions of eqs. (24) being 
{\it linear} in the antifields $\phi^*_{A a}$, $\bar{\phi}_A$ and $\eta_A$. 
Such solutions can be cast into the form (see eq. (18)) 
\begin{equation*}
S_m = S_{\rm cl} + 
(\hbox{$\frac{1}{2}$} \epsilon_{ab} \mathbf{s}_m^b \mathbf{s}_m^a + m^2) X,
\end{equation*}
with
$X = {\bar A}_i A^i + {\bar B}_{\alpha_0} B^{\alpha_0} +
{\bar C}_{\alpha_0 c} C^{\alpha_0 c}$. A realization of the (anti)BRST- and 
the $Sp(2)$-transformations of the antifields are given in Appendix E. 
Thus, we are left with the exercise to determine the corresponding 
transformations for the fields $A^i$, $B^{\alpha_0}$ and $C^{\alpha_0 c}$. 
Let us briefly look at the derivation of these transformations. Imposing the 
$osp(1,2)$-superalgebra 
\begin{equation}
[ \mathbf{d}_\alpha, \mathbf{d}_\beta ] = 
\epsilon_{\alpha\beta}^{~~~\!\gamma} \mathbf{d}_\gamma, 
\qquad
[ \mathbf{d}_\alpha, \mathbf{s}_m^a ] = 
\mathbf{s}_m^b (\sigma_\alpha)_b^{~a},
\qquad
\{ \mathbf{s}_m^a, \mathbf{s}_m^b \} = - 
m^2 (\sigma^\alpha)^{ab} \mathbf{d}_\alpha,
\end{equation}
on the gauge fields $A^i$, due to $\mathbf{d}_\alpha A^i = 0$, this yields 
$\{ \mathbf{s}_m^a, \mathbf{s}_m^b \} A^i = 0$. Then, with
\begin{equation}
\mathbf{s}_m^a A^i = R^i_{\alpha_0} C^{\alpha_0 a},
\end{equation}
by virtue of (50), we find
\begin{equation*}
R^i_{\alpha_0}
( \mathbf{s}_m^a C^{\alpha_0 b} + \mathbf{s}_m^b C^{\alpha_0 a} +
F^{\alpha_0}_{\beta_0 \gamma_0} C^{\beta_0 a} C^{\gamma_0 b} ) = 0.
\end{equation*}
Because the $R^i_{\alpha_0}$ are irreducible the general solution of this 
equation is given by
\begin{equation}
\mathbf{s}_m^a C^{\alpha_0 b} = \epsilon^{ab} B^{\alpha_0} -
\hbox{$\frac{1}{2}$}
F^{\alpha_0}_{\beta_0 \gamma_0} C^{\beta_0 a} C^{\gamma_0 b}.
\end{equation}
Imposing the superalgebra (52) on the (anti)ghost fields 
$C^{\alpha_0 c}$ and taking into account $\mathbf{d}_\alpha C^{\alpha_0 b} = 
C^{\alpha_0 c} (\sigma_\alpha)_c^{~b}$ it gives
$\{ \mathbf{s}_m^a, \mathbf{s}_m^b \} C^{\alpha_0 c} =  
- m^2 (\sigma^\alpha)^{ab} C^{\alpha_0 d} (\sigma_\alpha)_d^{~c}$.
The right-hand side of this restriction can be rewritten by means of the
relations $(\sigma_\alpha)_d^{~c} =
\epsilon_{de} \epsilon^{fc} (\sigma_\alpha)^e_{~f}$ and
$(\sigma^\alpha)^{ab} (\sigma_\alpha)^e_{~f} =
- (\epsilon^{ae} \delta^b_f + \epsilon^{be} \delta^a_f)$ as
$\{ \mathbf{s}_m^a, \mathbf{s}_m^b \} C^{\alpha_0 c} =  
- m^2 (\epsilon^{ac} C^{\alpha_0 b} + \epsilon^{bc} C^{\alpha_0 a})$.
Then, with (54), by virtue of (51), we obtain
\begin{align*}
\bigr\{&
\epsilon^{bc} (\mathbf{s}_m^a B^{\alpha_0} + m^2 C^{\alpha_0 a} -
\hbox{$\frac{1}{2}$} F^{\alpha_0}_{\beta_0 \gamma_0} 
B^{\beta_0} C^{\gamma_0 a} ) 
\\
& + \hbox{$\frac{1}{4}$} (
F^{\alpha_0}_{\eta_0 \beta_0} F^{\eta_0}_{\gamma_0 \delta_0} +
F^{\alpha_0}_{\eta_0 \delta_0} F^{\eta_0}_{\beta_0 \gamma_0} -
2 R^i_{\beta_0} F^{\alpha_0}_{\gamma_0 \delta_0, i} ) 
C^{\beta_0 a} C^{\gamma_0 b} C^{\delta_0 c} \bigr\} +
\hbox{sym}(a \leftrightarrow b) = 0,
\end{align*}
where $\hbox{sym}(a \leftrightarrow b)$ means symmetrization with respect 
to the indices $a$ and $b$.

The general solution of this equation reads
\begin{equation}
\mathbf{s}_m^a B^{\alpha_0} = - m^2 C^{\alpha_0 a} +
\hbox{$\frac{1}{2}$} 
F^{\alpha_0}_{\beta_0 \gamma_0} B^{\beta_0} C^{\gamma_0 a} +
\hbox{$\frac{1}{12}$} \epsilon_{cd} (
F^{\alpha_0}_{\eta_0 \beta_0} F^{\eta_0}_{\gamma_0 \delta_0} +
2 R^i_{\beta_0} F^{\alpha_0}_{\gamma_0 \delta_0, i} )
C^{\gamma_0 a} C^{\delta_0 c} C^{\beta_0 d}.
\end{equation}
For the particular case $m = 0$ the transformations (53)--(55) was already 
obtained earlier in Ref. \cite{12}.

Note that due to (52) the only non-zero variation of $A^i$ is of the form
\begin{equation}
\hbox{$\frac{1}{2}$} \epsilon_{ab} 
\mathbf{s}_m^b \mathbf{s}_m^a A^i = R^i_{\alpha_0} B^{\alpha_0} +
\hbox{$\frac{1}{2}$} \epsilon_{ab}  
R^i_{\alpha_0, j} R^j_{\beta_0} C^{\beta_0 b} C^{\alpha_0 a},
\end{equation}
for the corresponding variations of $C^{\alpha_0 c}$ and $B^{\alpha_0}$ one 
gets 
\begin{equation}
\hbox{$\frac{1}{2}$} \epsilon_{ab} 
\mathbf{s}_m^b \mathbf{s}_m^a C^{\alpha_0 c} = 
\hbox{$\frac{1}{2}$} m^2 C^{\alpha_0 c} - 
F^{\alpha_0}_{\beta_0 \gamma_0} B^{\beta_0} C^{\gamma_0 c} -
\hbox{$\frac{1}{6}$} \epsilon_{ab} (
F^{\alpha_0}_{\eta_0 \beta_0} F^{\eta_0}_{\gamma_0 \delta_0} +
2 R^i_{\beta_0} F^{\alpha_0}_{\gamma_0 \delta_0, i} )
C^{\gamma_0 c} C^{\delta_0 a} C^{\beta_0 b},
\end{equation}
\begin{equation}
\hbox{$\frac{1}{2}$} \epsilon_{ab} 
\mathbf{s}_m^b \mathbf{s}_m^a B^{\alpha_0} = - m^2 B^{\alpha_0}.
\end{equation}
The relations (53)--(58) specify the transformations of the 
$osp(1,2)$-symmetry for gauge theories with a closed algebra. Substituting 
these expressions into
\begin{align}
S_m &= S_{\rm cl} + A^*_{i a} (\mathbf{s}_m^a A^i) + 
B^*_{\alpha_0 a} (\mathbf{s}_m^a B^{\alpha_0}) -
C^*_{\alpha_0 a c} (\mathbf{s}_m^a C^{\alpha_0 c}) + 
( F_{\alpha_0 c} - \hbox{$\frac{1}{2}$} m^2 \bar{C}_{\alpha_0 c} ) 
C^{\alpha_0 c}
\nonumber\\
& \quad~ + {\bar A}_i (\hbox{$\frac{1}{2}$} 
\epsilon_{ab} \mathbf{s}_m^b \mathbf{s}_m^a A^i) +
{\bar B}_{\alpha_0} (\hbox{$\frac{1}{2}$} 
\epsilon_{ab} \mathbf{s}_m^b \mathbf{s}_m^a B^{\alpha_0}) +
\bar{C}_{\alpha_0 c} (\hbox{$\frac{1}{2}$} 
\epsilon_{ab} \mathbf{s}_m^b \mathbf{s}_m^a C^{\alpha_0 c}),
\end{align}
$F_{\alpha_0 c}$ being the only nonvanishing component of $\eta_A$; 
a direct verification shows that the resulting action $S_m$ satisfies 
eqs. (24) identically.

Finally, let us determine the gauge-fixed action $S_{m, {\rm eff}}$ in  
(36) for the class of {\it minimal} gauges $F$ depending only on $A^i$ and 
$C^{\alpha_0 c}$. Inserting into (42) for $S_m$ the action (59) and 
performing the integration over antifields and auxiliary fields we are led 
to the following expression for $S_{m, {\rm eff}}$ in the vacuum functional:  
\begin{equation*}
Z_m (0) = \int d A^i\, d B^{\alpha_0}\, d C^{\alpha_0 c}\,
{\exp}\bigr\{ (i/ \hbar) S_{m, {\rm eff}} \bigr\},
\qquad
S_{m, {\rm eff}} = S_{\rm cl} + W_F,
\end{equation*}
where $W_F$ is given by
\begin{align*}
W_F &= m^2 F + 
\hbox{$\frac{1}{2}$} \epsilon_{ab} (
\frac{\delta F}{\delta A^i}\, 
\mathbf{s}_m^b \mathbf{s}_m^a A^i +
\hbox{$\frac{1}{2}$} \epsilon_{ab} \frac{\delta F}{\delta C^{\alpha_0 c}}\,
\mathbf{s}_m^b \mathbf{s}_m^a C^{\alpha_0 c} )  
\\
& \quad - \hbox{$\frac{1}{2}$} \epsilon_{ab} ( 
\mathbf{s}_m^a A^i
\frac{\delta^2 F}{\delta A^i\, \delta A^j} 
\mathbf{s}_m^b A^j +
\mathbf{s}_m^a C^{\alpha_0 c}
\frac{\delta^2 F}{\delta C^{\alpha_0 c}\, \delta C^{\beta_0 d}}\, 
\mathbf{s}_m^b C^{\beta_0 d} ). 
\end{align*}
Again, this gauge-fixing term $W_F$ can we rewritten as
\begin{equation*}
W_F = (\hbox{$\frac{1}{2}$} \epsilon_{ab} 
\mathbf{s}_m^b \mathbf{s}_m^a + m^2) F,
\end{equation*}
showing that the action $S_{m, {\rm eff}}$ is in fact $osp(1,2)$-invariant 
and that the method of gauge fixing proposed in Sec. III will actually lift 
the degeneracy of the classical gauge-invariant action. As has been shown 
recently \cite{13}
in the particular case of Yang-Mills theories $S_{m, {\rm eff}}$ coincides 
with the gauge-fixed action in the massive Curci-Ferrari model \cite{14} in 
the Delbourgo-Jarvis gauge \cite{15}. The classical action $S_{\rm YM}$ is 
invariant under the non-abelian gauge transformations
\begin{equation*}
\delta A_\mu^\alpha = D_\mu^{\alpha \beta} \theta^\beta(x),
\qquad
D_\mu^{\alpha \beta} \equiv \delta^{\alpha \beta} \partial_\mu -
F^{\alpha \beta \gamma} A_\mu^\gamma,
\end{equation*}
where $F^{\alpha \beta \gamma}$ are the totally antisymmetric structure
constants. As before the $osp(1,2)$-invariance of the gauge-fixed action 
\begin{equation*}
S_{m, {\rm eff}} = S_{\rm YM} +
(\hbox{$\frac{1}{2}$} \epsilon_{ab} \mathbf{s}_m^b \mathbf{s}_m^a + m^2) F,
\qquad
F = \hbox{$\frac{1}{2}$} (A_\mu^\alpha A^{\mu \alpha} +
\xi \epsilon_{cd} C^{\alpha c} C^{\alpha d} ),
\end{equation*}
where $\xi$ is the gauge parameter, is assured by construction. Using
\begin{align*}
\mathbf{s}_m^a A_\mu^\alpha &= 
D_\mu^{\alpha \beta} C^{\beta a},
\\
\mathbf{s}_m^a C^{\alpha b} &= 
\epsilon^{ab} B^\alpha - \hbox{$\frac{1}{2}$} F^{\alpha \beta \gamma} 
C^{\beta a} C^{\gamma b},
\\
\mathbf{s}_m^a B^\alpha &= - m^2 C^{\alpha a} +
\hbox{$\frac{1}{2}$} F^{\alpha \beta \gamma} B^\beta C^{\gamma a} +
\hbox{$\frac{1}{12}$} \epsilon_{cd} 
F^{\alpha \eta \beta} F^{\eta \gamma \delta} 
C^{\gamma a} C^{\delta c} C^{\beta d},
\end{align*}
for the gauge-fixing terms one gets
\begin{align*}
\hbox{$\frac{1}{4}$} \epsilon_{ab} \mathbf{s}_m^b \mathbf{s}_m^a (
A_\mu^\alpha A^{\mu \alpha} ) &=
A_\mu^\alpha \partial^\mu B^\alpha + \hbox{$\frac{1}{2}$} \epsilon_{ab} (
\partial^\mu C^{\alpha b} ) D_\mu^{\alpha \beta} C^{\beta a},
\\
\hbox{$\frac{1}{4}$} \epsilon_{ab} \epsilon_{cd} 
\mathbf{s}_m^b \mathbf{s}_m^a (C^{\alpha c} C^{\alpha d}) &=
\hbox{$\frac{1}{2}$} m^2 \epsilon_{cd} C^{\alpha c} C^{\alpha d} +
B^\alpha B^\alpha - \hbox{$\frac{1}{24}$} \epsilon_{ab} \epsilon_{cd}
F^{\eta \alpha \beta} F^{\eta \gamma \delta} 
C^{\alpha a} C^{\beta c} C^{\gamma b} C^{\delta d}.
\end{align*}
The elimination of $B^\alpha$ can be performed by gaussian integration; it
provides the gauge-fixing term $\hbox{$\frac{1}{2}$} \xi^{-1} 
(\partial^\mu A_\mu^\alpha)^2$ and, in addition, among other interactions 
quartic (anti)ghost terms. This shows that the degeneracy of the classical 
action is indeed removed.

Concluding, we shall emphasize that, up to now, we have ignored the important 
question whether the action (59) is the general solution of the eqs. (24), 
i.e. being stable against perturbations. Unfortunately, this is not the case, 
because the fields $A^i$, $B^{\alpha_0}$, $C^{\alpha_0 c}$ and the antifields 
${\bar A}_i$, ${\bar B}_{\alpha_0}$, ${\bar C}_{\alpha_0 c}$ have the same 
quantum numbers and hence mix under renormalization. Therefore, we are 
confronted with the complicated problem how the action (59) must be changed 
in order to ensure the required stability. To attack this problem one is 
forced to introduce within (59) the antifields in a {\it nonlinear} manner 
(see Ref. \cite{13}); however, then the corresponding altered action cannot be 
expressed in such simple form as in (59). The solution of that problem will 
be given in another paper \cite{16}.
\bigskip\medskip
\begin{flushleft}
{\large{\bf VI. A Sp(2)-NONSYMMETRIC SOLUTION FOR $\mathbf{m = 0}$}} 
\end{flushleft}
\bigskip
We shall construct a solution $S(\zeta)$ of the master equations in the
BLT approach
\begin{equation}
\hbox{$\frac{1}{2}$} \bigr( S(\zeta), S(\zeta) \bigr)^a + V^a S(\zeta) = 0
\end{equation}
depending on an additional arbitrary parameter $\zeta$ (for the sake of 
simplicity we restrict ourselves to the case that $\zeta$ is 
field-independent) in such a way that the action is $Sp(2)$-nonsymmetric for 
$\zeta \neq 0$. Furthermore, we shall show that this solution $S(\zeta)$ 
exhibits an additional symmetry which may be used to single out the 
$Sp(2)$-symmetric solution.

To begin with, let us consider the $Sp(2)$-symmetric solution (see eq. (59))
\begin{align*}
S &= S_{\rm cl} + A^*_{i a} (\mathbf{s}^a A^i) + 
B^*_{\alpha_0 a} (\mathbf{s}^a B^{\alpha_0}) -
C^*_{\alpha_0 a c} (\mathbf{s}^a C^{\alpha_0 c}) 
\\
& \quad~ + {\bar A}_i ( \hbox{$\frac{1}{2}$} 
\epsilon_{ab} \mathbf{s}^b \mathbf{s}^a A^i ) +
{\bar B}_{\alpha_0} ( \hbox{$\frac{1}{2}$} 
\epsilon_{ab} \mathbf{s}^b \mathbf{s}^a B^{\alpha_0} ) +
\bar{C}_{\alpha_0 c} ( \hbox{$\frac{1}{2}$} 
\epsilon_{ab} \mathbf{s}^b \mathbf{s}^a C^{\alpha_0 c} ) 
\end{align*}
of the master equations (24) for $m = 0$ ($F_{\alpha_0 c}$ has been put equal 
to zero):
\begin{equation}
\hbox{$\frac{1}{2}$} (S,S)^a + V^a S = 0,
\end{equation}
where the (anti)BRST-transformations of $A^i$, $B^{\alpha_0}$ and 
$C^{\alpha_0 c}$ are given by (53)--(55). Note, that from (54) one gets the 
relation $B^{\alpha_0} = \hbox{$\frac{1}{2}$} \epsilon_{ab} 
\mathbf{s}^b C^{\alpha_0 a}$ which may be regarded as the {\it definition} of 
$B^{\alpha_0}$. 

Let us emphasize, that the (anti)BRST-transformations, as commonly used in the 
literature, contain no parameters. However, it is possible to incorporate into 
them a parameter $\zeta$ in a nontrivial way without violating the basic 
anticommutation relation $\{ \mathbf{s}^a, \mathbf{s}^b \} = 0$ and without 
changing the definition $B^{\alpha_0} = \hbox{$\frac{1}{2}$} 
\epsilon_{ab} \mathbf{s}^b C^{\alpha_0 a}$. To see this, we carry out in 
(53)--(55) the following nonlinear replacements: 
\begin{equation}
C^{\alpha_0 a} \rightarrow \chi_b^a(\zeta) C^{\alpha_0 b},
\qquad
B^{\alpha_0} \rightarrow B^{\alpha_0} - \hbox{$\frac{1}{4}$} 
\chi_{ab}(\zeta) F^{\alpha_0}_{\beta_0 \gamma_0}
C^{\beta_0 a} C^{\gamma_0 b},
\end{equation}
with $\chi_{ab}(\zeta) = \chi_a^c(\zeta) \epsilon_{cb}$, where the 
$Sp(2)$-nonsymmetric diagonal matrix $\chi_b^a(\zeta)$ is given by
\begin{equation*}
\chi_b^a(\zeta) \equiv (1 - \zeta^2)^{-1} \bigr(
\delta^a_b + \zeta (\sigma_0)^a_{~b} \bigr) =
\begin{pmatrix} (1 + \zeta)^{-1} & 0 \\ 0 & (1 - \zeta)^{-1} \end{pmatrix},
\qquad
\zeta \neq \pm 1.
\end{equation*}
Defining $\chi_c^a(\zeta) \bar{\chi}_b^c(\zeta) \equiv \delta^a_b$, with
$\bar{\chi}^{ab}(\zeta) = \epsilon^{ac} \bar{\chi}_c^b(\zeta)$ this leads to 
\begin{align*}
\mathbf{s}^a(\zeta) A^i &= R^i_{\alpha_0} \chi_b^a(\zeta) C^{\alpha_0 b},
\\
\mathbf{s}^a(\zeta) C^{\alpha_0 b} &= \bar{\chi}^{ab}(\zeta) B^{\alpha_0} -
\hbox{$\frac{1}{2}$} \bigr(
\chi_c^a(\zeta) \delta^b_d + 
\hbox{$\frac{1}{2}$} \bar{\chi}^{ab}(\zeta) \chi_{cd}(\zeta) \bigr)
F^{\alpha_0}_{\beta_0 \gamma_0} C^{\beta_0 c} C^{\gamma_0 d},
\\
\mathbf{s}^a(\zeta) B^{\alpha_0} &= \hbox{$\frac{1}{2}$} 
F^{\alpha_0}_{\beta_0 \gamma_0} B^{\beta_0} C^{\gamma_0 a} + 
\hbox{$\frac{1}{12}$} \epsilon_{bc} (
F^{\alpha_0}_{\eta_0 \beta_0} F^{\eta_0}_{\gamma_0 \delta_0} +
2 R^i_{\beta_0} F^{\alpha_0}_{\gamma_0 \delta_0, i} )
\chi^a_d(\zeta) C^{\gamma_0 d} C^{\delta_0 b} C^{\beta_0 c}.
\end{align*}
It is easy to prove that the (anti)BRST-operators $\mathbf{s}^a(\zeta)$ are 
nilpotent and anticommute among themselves as before, 
$\{ \mathbf{s}^a(\zeta), \mathbf{s}^b(\zeta) \} = 0$,
and it holds $B^{\alpha_0} = \hbox{$\frac{1}{2}$} \epsilon_{ab}
\mathbf{s}^b(\zeta) C^{\alpha_0 a}$. Therefore, the $\zeta$-dependent action
\begin{align}
S(\zeta) &= S_{\rm cl} + A^*_{i a} (\mathbf{s}^a(\zeta) A^i) + 
B^*_{\alpha_0 a} (\mathbf{s}^a(\zeta) B^{\alpha_0}) -
C^*_{\alpha_0 a c} (\mathbf{s}^a(\zeta) C^{\alpha_0 c}) 
\nonumber\\
& \quad~ + {\bar A}_i ( \hbox{$\frac{1}{2}$} 
\epsilon_{ab} \mathbf{s}^b(\zeta) \mathbf{s}^a(\zeta) A^i ) +
{\bar B}_{\alpha_0} ( \hbox{$\frac{1}{2}$} 
\epsilon_{ab} \mathbf{s}^b(\zeta) \mathbf{s}^a(\zeta) B^{\alpha_0} ) +
\bar{C}_{\alpha_0 c} ( \hbox{$\frac{1}{2}$} 
\epsilon_{ab} \mathbf{s}^b(\zeta) \mathbf{s}^a(\zeta) C^{\alpha_0 c} ) 
\end{align}
is a solution of the equation (60), but it violates the $Sp(2)$-symmetry for 
$\zeta \neq 0$. 

Let us emphasize that the parameter $\zeta$ cannot be restricted to zero  
without further (symmetry) requirements. This may be understood as follows.
In order to write down eq. (60) one has to introduce antifields coupled to 
the (anti)BRST-transformations of the fields. These transformations are 
defined in such a way that (60) does not depend on $\zeta$ explicitly.
Hence, $\zeta$ would correspond to a field renormalization if we 
could perform in equation (61), after carrying out the replacements (62), 
a redefinition of the antifields in such a way that the solution (63) is 
$\zeta$-independent. However, this is impossible due to the nonlinearity of
these replacements. Therefore, $\zeta$ is neither a normalization factor of
any parameter nor does it correspond to a field redefinition: it is free 
parameter of the theory if antifields are included (if antifields are 
{\it not} included the introduction of $\zeta$ will be irrelevant). This 
means that the requirement (61) alone is not sufficient to guarantee 
$Sp(2)$-symmetry of the theory, rather one has either to impose the additional 
symmetry requirement
\begin{equation*}
\mathbf{d}_\alpha S(\zeta) = 0,
\qquad
\mathbf{d}_\alpha \equiv \phi^A (\sigma_\alpha)_A^{~~\!B}
\frac{\delta}{\delta \phi^B} + V_\alpha,
\end{equation*} 
or to consider massive gauge theories and to impose the conditions (24).

Indeed, one can introduce symmetry operators 
$\mathbf{d}_\alpha(\zeta)$, whose action on $A^i$, $C^{\alpha_0 a}$ and
$B^{\alpha_0}$ are defined by
\begin{align*}
\mathbf{d}_\alpha(\zeta) A^i &= 0,
\\
\mathbf{d}_\alpha(\zeta) C^{\alpha_0 a} &=
\bar{\chi}^a_c \chi^d_e C^{\alpha_0 e} (\sigma_\alpha)_d^{~c},
\\
\mathbf{d}_\alpha(\zeta) B^{\alpha_0} &=
\hbox{$\frac{1}{4}$} \chi_{ab}(\zeta) F^{\alpha_0}_{\beta_0 \gamma_0} \bigr(
\bar{\chi}^a_c(\zeta) C^{\beta_0 e} C^{\gamma_0 b} +
\bar{\chi}^b_c(\zeta) C^{\beta_0 a} C^{\gamma_0 e} \bigr)
\chi^d_e(\zeta) (\sigma_\alpha)_d^{~c},
\end{align*}
which together with $\mathbf{s}^a(\zeta)$ obey the commutation relations
$[ \mathbf{d}_\alpha(\zeta), \mathbf{d}_\beta(\zeta) ] = 
\epsilon_{\alpha\beta}^{~~~\!\gamma} \mathbf{d}_\gamma(\zeta)$ and 
$[ \mathbf{d}_\alpha(\zeta), \mathbf{s}_m^a(\zeta) ] = 
\mathbf{s}_m^b(\zeta) (\sigma_\alpha)_b^{~a}$. This additional invariance 
property of $S(\zeta)$ may be used to fix $\zeta$ equal to zero, i.e.
$\zeta = 0$, and to single out the $Sp(2)$-symmetric solution $S$, by 
imposing the requirement that also the generating equations of the 
$Sp(2)$-symmetry do {\it not} depend on $\zeta$ explicitly.
\pagebreak
\bigskip\medskip
\begin{flushleft}
{\large{\bf VII. CONCLUDING REMARKS}} 
\end{flushleft}
\bigskip
We have proved the possibility of a consistent generalization of the 
$Sp(2)$ quantization scheme based on the orthosymplectic 
superalgebra $osp(1,2)$. Introducing mass terms into the theory, which do 
not break the extended BRST symmetry, the quantum master equations of the 
$Sp(2)$-symmetry must be satisfied in order to fulfil the corresponding 
equations of the extended BRST symmetry. To ensure $Sp(2)$-invariance of the 
theory also in the massless case then besides of the requirement of extended 
BRST symmetry the action $S$ must be subjected to the requirement of 
$Sp(2)$-symmetry explicitly.

Finally, let us give some remarks concerning the renormalization of the
theory. In Ref. \cite{3} this problem was discussed rather formally; thereby 
it had been accepted the standard hypothesis on the existence of a 
regularization respecting the Ward identities for the extended 
BRST symmetries.

On the other hand, there is a well established renormalization scheme, 
incidentally the most general one known until now, namely the BPHZL scheme 
\cite{17}, which does not need any regularization. 
Moreover, the renormalized quantum action principles, which sum up all 
properties of the renormalized perturbation series, were firstly established 
using the BPHZL renormalization scheme (they were rederived in different 
other schemes confirming their character as general theorems in 
renormalization theory, independent of the renormalization scheme). These 
theorems are extremely powerful and suffice for discussing the most useful 
identities among renormalized Green functions, such as Ward identities or 
their generalizations describing gauge invariance (e.g. the existence proofs 
for generating equations established in Ref. \cite{1,2} could be proven also 
by means of the action principles {\it without} any reference to a given 
renormalization and regularization scheme). 
Furthermore, within the BPHZL scheme it is possible 
to perform renormalization when we are concerned with {\it massless} 
theories. If massless field are involved, ultraviolet subtractions would
generate drastic spurious infrared divergences. Lowenstein and Zimmermann 
introduced a more involved subtraction procedure which is free of spurious
infrared divergences. In order to avoid these divergences one has 
to introduce a mass $m^2 (s - 1)^2$ with $s$ being an additional infrared 
subtraction parameter. Then, the disease that ultraviolet subtractions lead 
to nonintegrable infrared singularities is cured by performing ultraviolet
subtractions and extra infrared subtractions with respect to $(s - 1)$. 
It had been proven rigorously that $s = 1$ defines the renormalized massless 
theory; i.e. the introduction of the mass $m^2 (1 - s)^2$ leads not to a 
contradictory massive theory. Of cource, mass terms {\it softly} violate 
the gauge dependence of the $S$-matrix so that after performing 
renormalization one may take the massless limit $s = 1$.

Of course, this renormalization scheme could be applied also in the
$Sp(2)$-approach, but then the introduction of mass terms would violate
the extended BRST symmetries. On the other hand, using the $osp(1,2)$-approach 
it is possible to introduce mass terms (at the stage of gauge fixing) in such 
a way that the Ward identities of the extended BRST symmetries will  
{\it not} be violated, provided the Ward identity of the $Sp(2)$-symmetry 
is fulfilled to all orders of perturbation theory!
Besides, we can use the quantum action principles to get rigorous 
statements without any reference to a given renormalization 
and regularization scheme.

Obviously, the $osp(1,2)$-covariant frame of quantization can be formulated
within the superfield approach \cite{9} without much effort. 
On the other hand, the generalization of the 
formalism to the case of $L$-stage reducible theories analogous to \cite{2}
will be more involved.
\bigskip\medskip
\begin{flushleft}
{\large{\bf ACKNOWLEDGEMENT}} 
\end{flushleft}
\bigskip
The work of P. M. L. is partially supported by the Russian Foundation
for Basic Research (RFBR), project 96-02-16017, as well as the grant INTAS,
96-0308, and grant of Ministry of General and Professional Education of
Russian Federation in field of Basic Natural Sciences. He also thanks 
Graduate College ,,Quantum Field Theory`` at Leipzig University for 
hospitality and fruitful cooperation.


\appendix

\bigskip\medskip
\begin{flushleft}
{\large{\bf APPENDIX A. SUPERALGEBRA osp(1,2)}} 
\end{flushleft}
\bigskip
The (anti)commutation relations of the superalgebra $osp(1,2)$ in the 
Cartan-Weyl basis read (see Ref. \cite{18}):
\begin{gather}
[ L_0, L_\pm ] = \pm L_\pm,
\qquad
[ L_+, L_- ] = 2 L_0,
\\ 
[ L_0, R_\pm ] = \pm \hbox{$\frac{1}{2}$} R_\pm,
\qquad
[ L_\pm, R_\mp ] = - R_\pm,
\qquad 
[ L_\pm, R_\pm ] = 0,
\\
\{ R_\pm, R_\pm \} = \pm \hbox{$\frac{1}{2}$} L_\pm,
\qquad
\{ R_+, R_- \} = \hbox{$\frac{1}{2}$} L_0,
\end{gather}
and for the fundamental representation these generators are given by:
\begin{gather*}
L_0 = \begin{pmatrix} \hbox{$\frac{1}{2}$} & 0 & 0 \\ 
                      0 & - \hbox{$\frac{1}{2}$} & 0 \\
                      0 & 0 & 0 \end{pmatrix},
\qquad
L_+ = \begin{pmatrix} 0 & 1 & 0 \\
                      0 & 0 & 0 \\
                      0 & 0 & 0 \end{pmatrix},
\qquad
L_- = \begin{pmatrix} 0 & 0 & 0 \\
                      1 & 0 & 0 \\
                      0 & 0 & 0 \end{pmatrix},
\\
R_+ = \begin{pmatrix} 0 & 0 & \hbox{$\frac{1}{2}$} \\
                      0 & 0 & 0 \\
                      0 & \hbox{$\frac{1}{2}$} & 0 \end{pmatrix},
\qquad
R_- = \begin{pmatrix} 0 & 0 & 0 \\
                      0 & 0 & - \hbox{$\frac{1}{2}$} \\
                      \hbox{$\frac{1}{2}$} & 0 & 0 \end{pmatrix}.
\end{gather*}
The superalgebra (8)--(10) is obtained by the following identifications: 
\begin{equation*}
(\hbar/ i) \bar{\Delta}_0 = 2 L_0,
\qquad
(\hbar/ i) \bar{\Delta}_\pm = L_\pm,
\qquad
(\hbar/ i) \bar{\Delta}_m^1 = 2 m R_+,
\qquad
(\hbar/ i) \bar{\Delta}_m^2 = 2 m R_-;
\end{equation*}
so the relations (64)--(66) may be written instead as follows 
($\alpha = 0,+,-$):
\begin{align}
[ \bar{\Delta}_\alpha, \bar{\Delta}_\beta ] &= (i/ \hbar)
\epsilon_{\alpha\beta}^{~~~\!\gamma} \bar{\Delta}_\gamma, 
\\
[ \bar{\Delta}_\alpha, \bar{\Delta}_m^a ] &= (i/ \hbar) 
\bar{\Delta}_m^b (\sigma_\alpha)_b^{~a}, 
\\
\{ \bar{\Delta}_m^a, \bar{\Delta}_m^b \} &= - (i/ \hbar) m^2 
(\sigma^\alpha)^{ab} \bar{\Delta}_\alpha,
\end{align}
$\sigma_\alpha$ generating the group of special linear transformations with 
the $sl(2)$ algebra:
\begin{gather*}
\sigma_\alpha \sigma_\beta = g_{\alpha\beta} + 
\hbox{$\frac{1}{2}$} \epsilon_{\alpha\beta\gamma} \sigma^\gamma,
\qquad
\sigma^\alpha = g^{\alpha\beta} \sigma_\beta,
\qquad
\hbox{Tr}(\sigma_\alpha \sigma_\beta) = 2 g_{\alpha\beta},
\\
g^{\alpha\beta} = \begin{pmatrix} 1 & 0 & 0 \\  0 & 0 & 2 \\ 0 & 2 & 0 
\end{pmatrix},
\qquad
g^{\alpha\gamma} g_{\gamma\beta} = \delta^\alpha_\beta,
\end{gather*}
$\epsilon_{\alpha\beta\gamma}$ being the antisymmetric tensor, 
$\epsilon_{0+-} = 1$. Then, from (68) the following realization of 
$\sigma_\alpha$ is obtained
\begin{alignat*}{3}
(\sigma_+)_a^{~b} &= \begin{pmatrix} 0 & - 1 \\ 0 & 0 \end{pmatrix},
& \qquad
(\sigma_-)_a^{~b} &= \begin{pmatrix} 0 & 0 \\ - 1 & 0 \end{pmatrix},
& \qquad
(\sigma_0)_a^{~b} &= \begin{pmatrix} 1 & 0 \\ 0 & - 1 \end{pmatrix},
\\
\intertext{and by raising the first index according to
$(\sigma_\alpha)^{ab} = \epsilon^{ac} (\sigma_\alpha)_c^{~b}$ we get}
(\sigma_+)^{ab} &= \begin{pmatrix} 0 & 0 \\ 0 & 1 \end{pmatrix},
& \qquad
(\sigma_-)^{ab} &= \begin{pmatrix} - 1 & 0 \\ 0 & 0 \end{pmatrix},
& \qquad
(\sigma_0)^{ab} &= \begin{pmatrix} 0 & - 1 \\ - 1 & 0 \end{pmatrix}.
\end{alignat*}
\bigskip\medskip
\begin{flushleft}
{\large{\bf APPENDIX B. PROPERTIES OF THE BRACKETS}} 
\end{flushleft}
\bigskip
From the definitions (25), (26) and (29) it follows
\begin{alignat*}{2}
\epsilon(\{ F,G \}_\alpha) &= \epsilon(F) + \epsilon(G),
&\qquad
\{ F,G \}_\alpha &= \{ G,F \}_\alpha (-1)^{\epsilon(F) \epsilon(G)},
\\
\epsilon(( F,G )^a) &= \epsilon(F) + \epsilon(G) + 1,
&\qquad
( F,G )^a &= - ( G,F )^a (-1)^{(\epsilon(F) + 1)(\epsilon(G) + 1)},
\end{alignat*}
i.e. $\{ F,G \}_\alpha$ ($( F,G )^a$) defines an even (odd) graded bracket, 
and
\begin{align}
\Delta_\alpha (FG) &= (\Delta_\alpha F) G + F (\Delta_\alpha G) +
\{ F,G \}_\alpha,
\\
\Delta^a (FG) &= (\Delta^a F) G + F (\Delta^a G) (-1)^{\epsilon(F)} +
( F,G )^a (-1)^{\epsilon(F)},
\end{align}
where the last two relations may be understood as the definitions of the 
new brackets $\{ F,G \}_\alpha$ and the extended antibrackets
$( F,G )^a$ introduced in Ref. \cite{1}, respectively. Furthermore,
it holds
\begin{align*}
\{ F,GH \}_\alpha &= \{ F,G \}_\alpha H + G \{ F,H \}_\alpha
(-1)^{\epsilon(F) \epsilon(G)},
\\
( F,GH )^a &= ( F,G )^a H + G ( F,H )^a 
(-1)^{(\epsilon(F) + 1) \epsilon(G)},
\end{align*}
i.e. both brackets act on the algebra of functions under multiplications.

Let us now shortly state the properties of the new brackets
$\{ F,G \}_\alpha$ and $( F,G )^a$. Applying the following identities 
(their validity is verified by means of (29))
\begin{equation}
[ \Delta_\alpha, \Delta_\beta ] = 0,
\qquad
\{ \Delta^a, \Delta^b \} = 0,
\qquad
[ \Delta_\alpha, \Delta^a ] = 0,
\end{equation}
of the operators $\Delta^a$ and $\Delta_\alpha$ to a product of two functions 
$FG$ and making use of (70), (71), one gets
\begin{align*}
\Delta_{[\alpha} \{ F,G \}_{\beta]} &= \{ \Delta_{[\alpha} F,G \}_{\beta]} +
\{ F,\Delta_{[\alpha} G \}_{\beta]},
\\
\Delta^{\{a} ( F,G )^{b\}} &= ( \Delta^{\{a} F,G )^{b\}} +
( F,\Delta^{\{a} G )^{b\}} (-1)^{\epsilon(F) + 1},
\\
\Delta_\alpha ( F,G )^a - \Delta^a \{ F,G \}_\alpha (-1)^{\epsilon(F)} &=
( \Delta_\alpha F,G )^a + ( F, \Delta_\alpha G )^a
\\
& \quad~ - \{ \Delta^a F,G \}_\alpha (-1)^{\epsilon(F)} - 
\{ F, \Delta^a G \}_\alpha,
\end{align*}
where the square (curly) bracket indicates antisymmetrization (symmetrization)
in the indices $\alpha$ and $\beta$ ($a$ and $b$), respectively.  
Next, applying the relations (72) to a product of three functions $FGH$ by
means of simple but cumbersome calculations, one arrives at the following
Jacobi identities satisfied by two brackets:
\begin{align}
\{ \{ F,G \}_{[\alpha},H \}_{\beta]} (-1)^{\epsilon(F) \epsilon(H)} +
\hbox{cyclic}(F,G,H) &\equiv 0,
\\
( ( F,G )^{\{a},H )^{b\}} (-1)^{(\epsilon(F) + 1)(\epsilon(H) + 1)} +
\hbox{cyclic}(F,G,H) &\equiv 0,
\\
\bigr(
\{ ( F,G )^a,H \}_\alpha - ( \{ F,G \}_\alpha,H )^a (-1)^{\epsilon(G)} \bigr) 
(-1)^{\epsilon(F) (\epsilon(H) + 1)} +
\hbox{cyclic}(F,G,H) &\equiv 0.
\end{align}
Furthermore, one needs the operators $V_\alpha$ and $V_m^a$, 
eqs. (27) and (28), to obey the $osp(1,2)$-superalgebra:
\begin{equation}
[ V_\alpha, V_\beta ] = \epsilon_{\alpha\beta}^{~~~\!\gamma} V_\gamma,
\qquad
\{ V_m^a, V_m^b \} = - m^2 (\sigma^\alpha)^{ab} V_\alpha,
\qquad
[ V_\alpha, V_m^a ] = V_m^b (\sigma_\alpha)_b^{~a}.
\end{equation}
Applying the identities (their validity is established by means of 
the eqs. (27)--(29))
\begin{align}
[ \Delta_\alpha, V_\beta ] + [ V_\alpha, \Delta_\beta ] &= 
\epsilon_{\alpha\beta}^{~~~\!\gamma} \Delta_\gamma,
\nonumber\\
\{ \Delta^a, V_m^b \} + \{ V_m^a, \Delta^b \} &= 
- m^2 (\sigma^\alpha)^{ab} \Delta_\alpha,
\\
[ \Delta_\alpha, V_m^a ] + [ V_\alpha, \Delta^a ] &= 
\Delta^b (\sigma_\alpha)_b^{~a},
\nonumber
\end{align}
to a product of two functions $FG$, one can establish the following relations 
which define the action of the operators $V_\alpha$ and $V_m^a$ upon the 
brackets,
\begin{align*}
V_{[\alpha} \{ F,G \}_{\beta]} &= 
\epsilon_{\alpha\beta}^{~~~\!\gamma} \{ F,G \}_\gamma +
\{ V_{[\alpha} F,G \}_{\beta]} +
\{ F,V_{[\alpha} G \}_{\beta]},
\\
V_m^{\{a} ( F,G )^{b\}} &= - m^2 (\sigma^\alpha)^{ab} \{ F,G \}_\alpha +
( V_m^{\{a} F,G )^{b\}} + ( F,V_m^{\{a} G )^{b\}} (-1)^{\epsilon(F) + 1},
\\
V_\alpha ( F,G )^a &- V_m^a \{ F,G \}_\alpha (-1)^{\epsilon(F)} = 
( F,G )^b (\sigma_\alpha)_b^{~a} +
( V_\alpha F,G )^a + ( F,V_\alpha G )^a
\\
& \qquad\qquad\qquad\qquad\qquad\quad
- \{ V_m^a F,G \}_\alpha (-1)^{\epsilon(F)} - \{ F,V_m^a G \}_\alpha,
\end{align*}
where, as before, square (curly) bracket means antisymmetrization 
(symmetrization) in the indices $\alpha$ and $\beta$ ($a$ and $b$), 
respectively. 

Note that the properties (76) of $V_m^a$ and $V_\alpha$ are inherited by the
the operators $\bar{\Delta}_m^a = \Delta^a + (i/ \hbar) V_m^a$ and 
$\bar{\Delta}_\alpha = \Delta_\alpha + (i/ \hbar) V_\alpha$, 
see eqs. (8)--(10), due to the properties (72) and (77). 

Finally, let us specify $F = G = H \equiv S$ to be any bosonic functional $S$,
$\epsilon(S) = 0$. Then, the Jacobi identities simplify into
\begin{equation}
\{ \{ S,S \}_{[\alpha},S \}_{\beta]} \equiv 0,
\qquad
( ( S,S )^{\{a},S )^{b\}} \equiv 0,
\qquad
\{ ( S,S )^a,S \}_\alpha - ( \{ S,S \}_\alpha,S )^a \equiv 0.
\end{equation}
For the action of the operators $\Delta_\alpha$, $\Delta^a$ and
$V_\alpha$, $V_m^a$ upon the brackets $\{ S,S \}_\alpha$ and
$( S,S )^a$ one becomes:
\begin{align}
\hbox{$\frac{1}{2}$} \Delta_{[\alpha} \{ S,S \}_{\beta]} &= 
\{ \Delta_{[\alpha} S,S \}_{\beta]},  
\nonumber\\
\hbox{$\frac{1}{2}$} \Delta^{\{a} ( S,S )^{b\}} &= 
( \Delta^{\{a} S,S )^{b\}},
\\
\hbox{$\frac{1}{2}$} \bigr( \Delta_\alpha ( S,S )^a - 
\Delta^a \{ S,S \}_\alpha \bigr) &=
( \Delta_\alpha S,S )^a - \{ \Delta^a S,S \}_\alpha
\nonumber\\
\intertext{and}
\hbox{$\frac{1}{2}$} V_{[\alpha} \{ S,S \}_{\beta]} &= 
\{ V_{[\alpha} S,S \}_{\beta]} +
\hbox{$\frac{1}{2}$} \epsilon_{\alpha\beta}^{~~~\!\gamma} \{ S,S \}_\gamma,
\nonumber
\\
\hbox{$\frac{1}{2}$} V_m^{\{a} ( S,S )^{b\}} &= 
( V_m^{\{a} S,S )^{b\}} - 
\hbox{$\frac{1}{2}$} m^2 (\sigma^\alpha)^{ab} \{ S,S \}_\alpha,
\\
\hbox{$\frac{1}{2}$} \bigr( V_\alpha ( S,S )^a -
V_m^a \{ S,S \}_\alpha \bigr) &= 
( V_\alpha S,S )^a - \{ V_m^a S,S \}_\alpha +
\hbox{$\frac{1}{2}$} ( S,S )^b (\sigma_\alpha)_b^{~a}.
\nonumber
\end{align}
\bigskip\medskip
\begin{flushleft}
{\large{\bf APPENDIX C. COMPATIBILITY PROOF  
\\
$\phantom{\bf APPENDIX C.~}$ FOR GENERATING EQUATIONS}} 
\end{flushleft}
\bigskip
The existence of a solution, now denoted by $S_{m(0)}$, of the generating 
equations at the lowest order approximation has been verified in (23) and
(24), 
\begin{equation}
\hbox{$\frac{1}{2}$} ( S_{m(0)}, S_{m(0)} )^a + V_m^a S_{m(0)} = 0,
\qquad
\hbox{$\frac{1}{2}$} \{ S_{m(0)}, S_{m(0)} \}_\alpha + V_\alpha S_{m(0)} = 0.
\end{equation}
Let us prove that the generalization of these equations at any order of
$\hbar$, namely
\begin{equation}
\hbox{$\frac{1}{2}$} ( S_m, S_m )^a + V_m^a S_m +
(\hbar/ i) \Delta^a S_m = 0,
\qquad
\hbox{$\frac{1}{2}$} \{ S_m, S_m \}_\alpha + V_\alpha S_m +
(\hbar/ i) \Delta_\alpha S_m = 0,
\end{equation}
for the full quantum action
\begin{equation*}
S_m = \sum_{n = 0}^\infty \hbar^n S_{m(n)}
\end{equation*}
is compatible with the algebraic properties of the antibrackets
$\{ S_m,S_m \}_\alpha$ and $( S_m,S_m )^a$ (see Appendix B).

Suppose that functionals $S_{m(k)}$, $k \leq n$, have been found
obeying eqs. (82) up to and including the $n$th order in $\hbar$. Then
for the $(n + 1)$th order the following equations should be satisfied:
\begin{equation}
Q^a_m S_{m(n + 1)} = Y^a_{m(n + 1)},
\qquad
Q_\alpha S_{m(n + 1)} = Y_{\alpha(n + 1)}, 
\end{equation}
where the operators $Q_m^a$ and $Q_\alpha$ are defined by
\begin{equation}
Q_m^a X \equiv ( S_{m(0)},X )^a + V_m^a X,
\qquad
Q_\alpha X \equiv \{ S_{m(0)}, X \}_\alpha + V_\alpha X,
\end{equation}
$X$ being an arbitrary functional. By virtue of eqs. (81) the operators
$Q_m^a$ and $Q_\alpha$ obey the $osp(1,2)$-superalgebra:
\begin{equation}
[ Q_\alpha,Q_\beta ] = \epsilon_{\alpha\beta}^{~~~\!\gamma} Q_\gamma,
\qquad
[ Q_\alpha,Q_m^a ] = Q_m^b (\sigma_\alpha)_b^{~a},
\qquad
\{ Q_m^a,Q_m^b \} = - m^2 (\sigma^\alpha)^{ab} Q_\alpha.
\end{equation}
The functionals $Y^a_{m(n + 1)}$ and $Y_{\alpha(n + 1)}$ in eqs. (83) have 
the following form
\begin{align*}
Y^a_{m(n + 1)} &= i \Delta^a S_{m(n)} - 
\hbox{$\frac{1}{2}$} \sum_{k = 1}^n ( S_{m(n + 1 - k)},S_{m(k)} )^a,
\\
Y_{\alpha(n + 1)} &= i \Delta_\alpha S_{m(n)} -
\hbox{$\frac{1}{2}$} \sum_{k = 1}^n \{ S_{m(n + 1 - k)},S_{m(k)} \}_\alpha.
\end{align*}

In order to ensure that the eqs. (83) are compatible with the 
$osp(1,2)$-superalgebra (85) it is necessary that the following relations
hold:
\begin{align}
Q_\alpha Y_{\beta(n + 1)} - Q_\beta Y_{\alpha(n + 1)} &= 
\epsilon_{\alpha\beta}^{~~~\!\gamma} Y_{\gamma(n + 1)},
\nonumber\\
Q^a_m Y^b_{m(n + 1)} + Q^b_m Y^a_{m(n + 1)} &= 
- m^2 (\sigma^\alpha)^{ab} Y_{\alpha(n + 1)},
\\
Q_\alpha Y^a_{m(n + 1)} - Q^a_m Y_{\alpha(n + 1)} &=
Y^b_{m(n + 1)} (\sigma_\alpha)_b^{~a}.
\nonumber
\end{align}
Let us show that the relations (86) indeed are satisfied. To begin with,
we consider the Jacobi identity 
$\{ S_m, \{ S_m,S_m \}_{[\alpha} \}_{\beta]} \equiv 0$, eqs. (78). By virtue 
of the algebraic properties (79) and (80) of $\{ S_m,S_m \}_\alpha$ the 
left-hand side of this identity can be rewritten as 
\begin{align}
\{ S_m, \hbox{$\frac{1}{2}$} \{ S_m,S_m \}_{[\alpha} \}_{\beta]} &\equiv 
\{ S_m, \hbox{$\frac{1}{2}$} \{ S_m,S_m \}_{[\alpha} +
V_{[\alpha} S_m + (\hbar/ i) \Delta_{[\alpha} S_m \}_{\beta]}
\nonumber\\
& \quad~ - 
\{ S_m, V_{[\alpha} S_m + (\hbar/ i) \Delta_{[\alpha} S_m \}_{\beta]} 
\nonumber\\
& = \{ S_m, \hbox{$\frac{1}{2}$} \{ S_m,S_m \}_{[\alpha} +
V_{[\alpha} S_m + (\hbar/ i) \Delta_{[\alpha} S_m \}_{\beta]}
\\
& \quad~ - V_{[\alpha} \bigr(
\hbox{$\frac{1}{2}$} \{ S_m,S_m \}_{\beta]} +
V_{\beta]} S_m + (\hbar/ i) \Delta_{\beta]} S_m \bigr)
\nonumber\\
& \quad~ - (\hbar/ i) \Delta_{[\alpha} \bigr(
\hbox{$\frac{1}{2}$} \{ S_m,S_m \}_{\beta]} +
V_{\beta]} S_m + (\hbar/ i) \Delta_{\beta]} S_m \bigr)
\nonumber\\
& \quad~ + \epsilon_{\alpha\beta}^{~~~\!\gamma} \bigr(
\hbox{$\frac{1}{2}$} \{ S_m,S_m \}_\gamma +
V_\gamma S_m + (\hbar/ i) \Delta_\gamma S_m \bigr).
\nonumber
\end{align}
Considering the generating equation (82) in the $(n + 1)$th order in $\hbar$:
\begin{equation*}
\hbox{$\frac{1}{2}$} \{ S_m, S_m \}_\alpha + V_\alpha S_m +
(\hbar/ i) \Delta_\alpha S_m = \hbar^{n + 1} (
Q_\alpha S_{m(n + 1)} - Y_{\alpha(n + 1)} ) + O(\hbar^{n + 2}),
\end{equation*}
then for the equality (87) in the $(n + 1)$th order one gets
\begin{equation}
Q_{[\beta} (Q_{\alpha]} S_{m(n + 1)} - Y_{\alpha](n + 1)} ) +
\epsilon_{\alpha\beta}^{~~~\!\gamma} (
Q_\gamma S_{m(n + 1)} - Y_{\gamma(n + 1)} ) = 0,
\end{equation}
where the definition (84) of the operator $Q_\alpha$ has been used.
Taking into account the properties (85) of $Q_\alpha$ we conclude from (88)
that the first relation (86) is fulfilled. In the same way, using the
Jacobi identities $( S_m, ( S_m,S_m )^{\{a} )^{b\}} \equiv 0$ and
$\{ S_m, ( S_m,S_m )^a \}_\alpha - ( S_m, \{ S_m,S_m \}_\alpha )^a \equiv 0$,
eqs. (78), it can be verified that also the remaining relations (86) are 
fulfilled; thus establishing the compatibility of eqs. (83). Obviously, 
by virtue of (85), the equations (86) are solved by 
\begin{equation}
Y^a_{m(n + 1)} = Q^a_m X_{m(n + 1)},
\qquad
Y_{\alpha(n + 1)} = Q_\alpha X_{m(n + 1)},
\end{equation}
with arbitrary functional $X_{m(n + 1)}$. Here, we do not check that {\it any} 
solution of (86) has the form (89), which means that the theory is anomaly 
free; in order to check anomaly freedom one can adopt the methods of 
Ref. \cite{1,2}. Choosing $S_{m(n + 1)} = X_{m(n + 1)}$ the 
generating equations (82) are satisfied up to and including the $(n + 1)$th 
order in $\hbar$ and we can proceed by induction.     

\bigskip\medskip
\begin{flushleft}
{\large{\bf APPENDIX D. COMPONENTWISE NOTATION   
\\
$\phantom{\bf APPENDIX D.~}$ OF THE TRANSFORMATIONS (19)}} 
\end{flushleft}
\bigskip
In componentwise notation the extended BRST- and $Sp(2)$-transformations (19) 
of the antifields read as follows ($D_i$ and $E_{\alpha_0}$ have been put
equal to zero):
\begin{alignat}{2}
V_m^a \bar{A}_i &= \epsilon^{ab} A_{i b}^*,
&\qquad
V_\alpha \bar{A}_i &= 0,
\nonumber\\
V_m^a A_{i b}^* &= m^2 \delta^a_b \bar{A}_i,
&\qquad
V_\alpha A_{i b}^* &= A_{i d}^* (\sigma_\alpha)^d_{~b},
\nonumber\\
V_m^a \bar{B}_{\alpha_0} &= \epsilon^{ab} B_{\alpha_0 b}^*,
&\qquad
V_\alpha \bar{B}_{\alpha_0} &= 0,
\nonumber\\
V_m^a B_{\alpha_0 b}^* &= m^2 \delta^a_b \bar{B}_{\alpha_0},
&\qquad
V_\alpha B_{\alpha_0 b}^* &= B_{\alpha_0 d}^* (\sigma_\alpha)^d_{~b},
\\
V_m^a \bar{C}_{\alpha_0 c} &= \epsilon^{ab} C_{\alpha_0 bc}^*,
&\qquad
V_\alpha \bar{C}_{\alpha_0 c} &= 
\bar{C}_{\alpha_0 d} (\sigma_\alpha)^d_{~c},
\nonumber\\
V_m^a F_{\alpha_0 c} &= m^2 \epsilon^{ab} ( 
C_{\alpha_0 bc}^* - C_{\alpha_0 cb}^* ),
&\qquad
V_\alpha F_{\alpha_0 c} &= F_{\alpha_0 d} (\sigma_\alpha)^d_{~c},
\nonumber\\
V_m^a C_{\alpha_0 bc}^* &= m^2 ( 
\delta^a_b \bar{C}_{\alpha_0 c} + \delta^a_c \bar{C}_{\alpha_0 b} ) - 
\delta^a_b F_{\alpha_0 c},
&\qquad
V_\alpha C_{\alpha_0 bc}^* &= C_{\alpha_0 dc}^* (\sigma_\alpha)^d_{~b} + 
C_{\alpha_0 bd}^* (\sigma_\alpha)^d_{~c},
\nonumber
\end{alignat}
where the additional antifield $F_{\alpha_0 c}$ has to be introduced in 
order to obey the $osp(1,2)$-superalgebra (76) (see Appendix C). 
Let us stress that expressing this algebra through operator identities is a 
stronger restriction than satisfying this algebra by means of the (anti)BRST 
transformations (90) which could be also realized without introducing 
$F_{\alpha_0 c}$, namely by choosing $C_{\alpha_0 ab}^* = C_{\alpha_0 ba}^*$.  

The componentwise notation of the operators $V_m^a$, $V_\alpha$ and 
$\Delta^a$, $\Delta_\alpha$ are given by (see (27)--(29))
\begin{align*}
V_m^a &= \epsilon^{ab} A_{i b}^* \frac{\delta}{\delta \bar{A}_i} +
m^2 \bar{A}_i \frac{\delta}{\delta A_{i a}^*} +
\epsilon^{ab} B_{\alpha_0 b}^* \frac{\delta}{\delta \bar{B}^{\alpha_0}} +
m^2 \bar{B}^{\alpha_0} \frac{\delta}{\delta B_{\alpha_0 b}^*} +
\epsilon^{ab} C_{\alpha_0 bc}^* 
\frac{\delta}{\delta \bar{C}_{\alpha_0 c}}
\\
& \quad~ + m^2 \bar{C}_{\alpha_0 c} ( 
\frac{\delta}{\delta C_{\alpha_0 ac}^*} +
\frac{\delta}{\delta C_{\alpha_0 ca}^*} ) -
F_{\alpha_0 c} \frac{\delta}{\delta C_{\alpha_0 ac}^*} +
m^2 \epsilon^{ab} ( 
C_{\alpha_0 bc}^* - C_{\alpha_0 cb}^* ) 
\frac{\delta}{\delta F_{\alpha_0 c}},
\\
V_\alpha &= A_{i d}^* (\sigma_\alpha)^d_{~b} 
\frac{\delta}{\delta A_{i b}^*} +
B_{\alpha_0 d}^* (\sigma_\alpha)^d_{~b}
\frac{\delta}{\delta B_{\alpha_0 b}^*} +
\bar{C}_{\alpha_0 d} (\sigma_\alpha)^d_{~c}
\frac{\delta}{\delta \bar{C}_{\alpha_0 c}}
\\
& \quad~ + \bigr(
C_{\alpha_0 dc}^* (\sigma_\alpha)^d_{~b} +
C_{\alpha_0 bd}^* (\sigma_\alpha)^d_{~c} \bigr)
\frac{\delta}{\delta C_{\alpha_0 bc}^*} +
F_{\alpha_0 d} (\sigma_\alpha)^d_{~c}
\frac{\delta}{\delta F_{\alpha_0 c}}
\\
\intertext{and}
\Delta^a &= (-1)^{\epsilon_i} \frac{\delta_L}{\delta A^i} 
\frac{\delta}{\delta A_{i a}^*} +
(-1)^{\epsilon_{\alpha_0}} \frac{\delta_L}{\delta B^{\alpha_0}} 
\frac{\delta}{\delta B_{\alpha_0 a}^*} +
(-1)^{\epsilon_{\alpha_0} + 1} \frac{\delta_L}{\delta C^{\alpha_0 c}} 
\frac{\delta}{\delta C_{\alpha_0 ac}^*}, 
\\
\Delta_\alpha &= (-1)^{\epsilon_{\alpha_0} + 1} 
(\sigma_\alpha)_d^{~c}
\frac{\delta_L}{\delta C^{\alpha_0 c}}
\frac{\delta}{\delta F_{\alpha_0 d}}. 
\end{align*}
Then, it can be verified that $V_m^a$, $V_\alpha$ and 
$\Delta^a$, $\Delta_\alpha$ fulfil the (anti)commutation relations 
(72), (76) and (77) (see Appendix C).


\end{document}